\documentclass{optica-article}

\journal{opticajournal} % for journals or Optica Open

\articletype{Research Article}

\usepackage{lineno}
\begin{document}

% \title{High-spatiotemporal-throughput microscopy via diffractive multiplexing across a gigapixel sensor array}
%\title{Highly parallelized, diffractively multiplexed computational microscopy}
%\title{Highly parallelized diffractive multiplexed computational microscopy}
%\title{Massively parallelized, diffractively multiplexed, compressive microscopy}
\title{Large-scale compressive microscopy via diffractive multiplexing across a sensor array}
% \title{Compressive microscopy with diffractive multiplexing and massively parallelized detection}
%\title{Multiplexed, parallelized, compressive microscopy}
%\title{Multiplexed parallel microscopy}
%\title{Diffractive multiplexing across a highly parallelized microscope}

\author{Kevin C. Zhou,\authormark{1,6}* Chaoying Gu,\authormark{1} Muneki Ikeda,\authormark{2} Tina M. Hayward,\authormark{3} Nicholas Antipa,\authormark{4} Rajesh Menon,\authormark{3} Roarke Horstmeyer,\authormark{5} Saul Kato,\authormark{2} and Laura Waller\authormark{1,}*}

\address{\authormark{1}Department of Electrical Engineering and Computer Sciences, University of California, Berkeley, CA\\
\authormark{2}Department of Neurology, University of California, San Francisco, CA\\
\authormark{3}Department of Electrical and Computer Engineering, University of Utah, Salt Lake City, UT\\
\authormark{4}Department of Electrical and Computer Engineering, University of California, San Diego, CA\\
\authormark{5}Department of Biomedical Engineering, Duke University, Durham, NC\\
\authormark{6}Department of Biomedical Engineering, University of Michigan, Ann Arbor, MI\\
}

\email{\authormark{*}kczhou@umich.edu, waller@berkeley.edu}

% use {asbstract*} to suppress the copyright line. Copyright information will be added in production

\begin{abstract*} 
Microscopes face a trade-off between spatial resolution, field-of-view, and frame rate -- improving one of these properties typically requires sacrificing the others, due to the limited spatiotemporal throughput of the sensor. To overcome this, we propose a new microscope that achieves snapshot gigapixel-scale imaging with a sensor array and a diffractive optical element (DOE). We improve the spatiotemporal throughput in two ways. First, we capture data with an array of 48 sensors resulting in 48$\times$ more pixels than a single sensor. Second, we use point spread function (PSF) engineering and compressive sensing algorithms to fill in the missing information from the gaps surrounding the individual sensors in the array, further increasing the spatiotemporal throughput of the system by an additional >5.4$\times$. The array of sensors is modeled as a single large-format ``super-sensor,'' with erasures corresponding to the gaps between the individual sensors. The array is placed at the output of a (nearly) 4f imaging system, and we design a DOE for the Fourier plane that generates a distributed PSF that encodes information from the entire super-sensor area, including the gaps. We then computationally recover the large-scale image, assuming the object is sparse in some domain. Our calibration-free microscope can achieve $\sim$3 \textmu m resolution over >5.2 cm\textsuperscript{2} FOVs at up to 120 fps, culminating in a total spatiotemporal throughput of 25.2 billion pixels per second. We demonstrate the versatility of our microscope in two different modes: structural imaging via darkfield contrast and functional fluorescence imaging of calcium dynamics across dozens of freely moving \textit{C. elegans} simultaneously.
%In particular, our computational microscope features an optimized diffractive pupil that multiplexes 
%In particular, our computational microscope employs a discontiguous array of 48 camera sensors, with the information in the gaps multiplexed via pupil coding with an optimized diffractive optical element.

%calibration free

%>5.4x fov/throughput enhancement

%compatible with darkfield and fluorescence imaging

%<3um res, >5.2cm2 FOV, 25 GP/sec

\end{abstract*}

%%%%%%%%%%%%%%%%%%%%%%%%%%  body  %%%%%%%%%%%%%%%%%%%%%%%%%%
\section{Introduction}
Optical engineers face tradeoffs that limit the spatiotemporal throughput of the microscopes they design. Strategies that expand the space-bandwidth product (SBP) -- the number of spatially resolvable points -- \cite{park2021review} typically sacrifice frame rate. For example, physically scanning the sample ~\cite{hanna2020whole} or using scanning mirrors \cite{potsaid2005adaptive,shi2024random} to build up a larger composite field-of-view (FOV) is time-consuming. Other techniques keep a constant FOV and improve resolution by capturing multiple images then computationally combining them; for example Fourier ptychography~\cite{zheng2013wide,konda2020fourier,zheng2021concept}, single-molecule localization microscopy \cite{lelek2021single}, and structured illumination microscopy (SIM) \cite{gustafsson2000surpassing}. These methods can increase SBP and can achieve faster acquisition times than sample scanning since they do not require any moving parts. However, any sequential measurements mean that the sample should remain static during the acquisition, limiting use for imaging dynamic samples. %In fact, even though these techniques increase SBP, the overall spatiotemporal throughput (SBP/time) may be the same or even reduced relative to standard widefield microscopy, due to oversampling requirements (e.g. FOV overlap in scanning microscopy, Fourier overlap in Fourier ptychography). 

To achieve faster imaging, multiplexing and compressive sensing strategies can enable fewer measurements, for example, by illuminating the sample with multiple LEDs simultaneously in Fourier ptychography~\cite{tian2014multiplexed, tian2015computational} or overlapping multiple FOVs \cite{malik2023single, horisaki2010multi, shepard2015design,yao2022increasing}, which are demixed computationally. More recently, computational approaches can model and correct for sample motion during acquisition~\cite{cao2024neural,kellman2018motion}, circumventing the need for static samples. However, for truly single-shot large spatiotemporal throughput imaging, multi-camera array architectures have proven most useful for enabling high-speed large-SBP video acquisition~\cite{fan2019video,zhou2023parallelized,harfouche2023imaging}. 

\begin{figure}
    \centering
    \includegraphics[width=\textwidth]{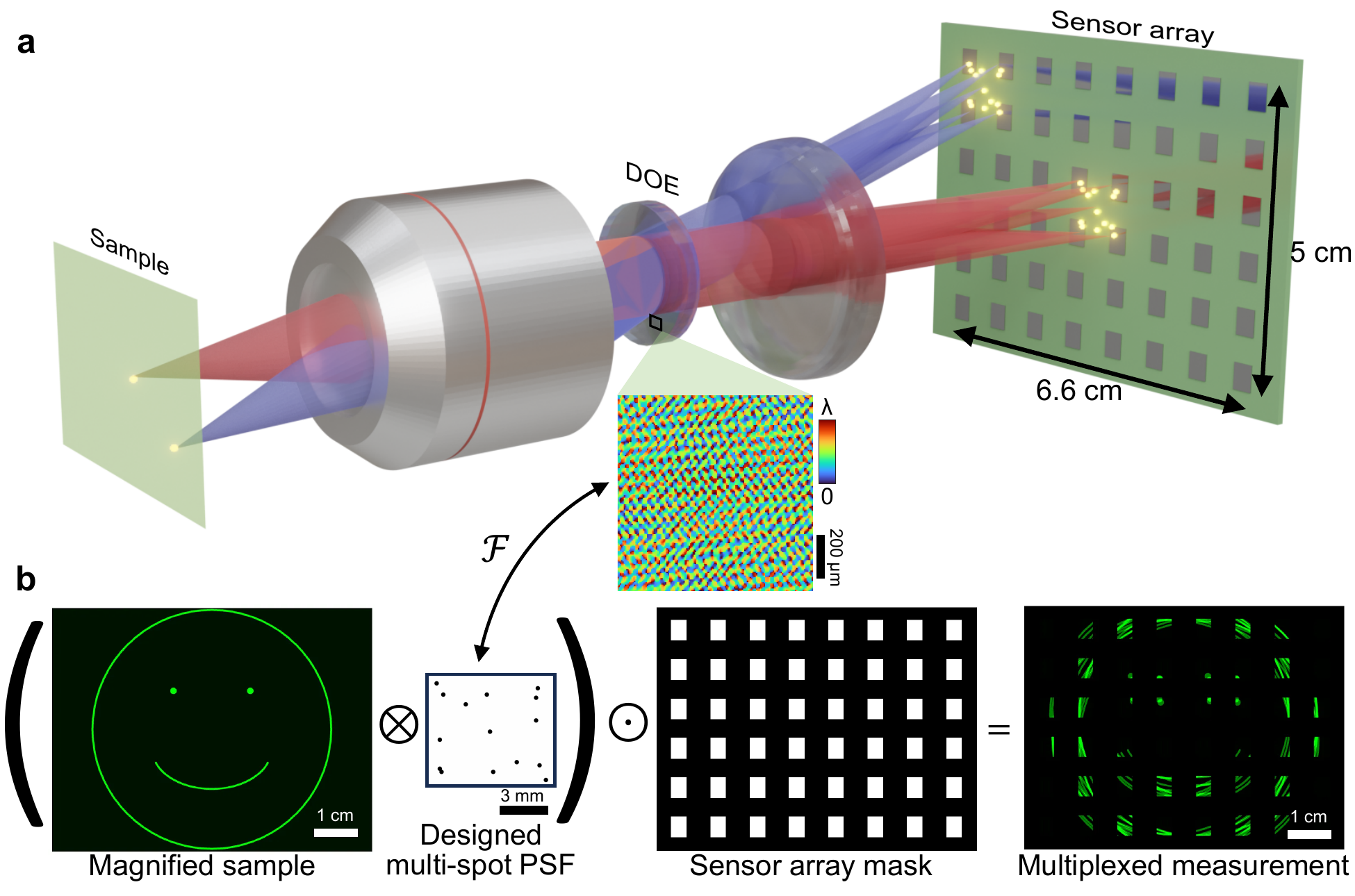}
    \caption{Overview of our large-scale compressive microscope. (a) The imaging system uses an engineered diffractive optical element (DOE) in the pupil plane to generate a distributed multi-spot point spread function (PSF) at the image plane. Images are captured by an array of 48 sensors arranged in a grid with gaps between individual sensors. (b) The system forward model is a masked convolution with the PSF, which compressively encodes the parts of the image that would normally fall on the gaps between the sensors. A computational inverse problem then recovers the full-scale image, including areas outside the sensor coverage.}
    \label{fig:teaser}
\end{figure}

Here, we present a new computational microscope that starts with a multi-sensor array and further improves SBP by spatial multiplexing to achieve single-shot gigapixel-scale imaging (Fig. \ref{fig:teaser}). In contrast to previous multi-camera systems, ours treats the sensor array as a single large-format sensor, without associating each individual sensor with its own lens. This keeps the optical design simple. To fill in the inter-sensor gaps and solve for the full-scale image, using only the data captured at the sensors (comprising only $\sim$22\% of the total area), we optimized and fabricated a diffractive optical element (DOE) to create a custom PSF that generates 15 shifted copies of the image, such that points that would otherwise have imaged onto the gaps are detected by at least one sensor in the array. Including both inter-sensor interpolation and extra-sensor extrapolation, our method increases the effective FOV by >5.4$\times$. Conveniently, our design does not require a physical PSF calibration step and is robust to system perturbations. We demonstrate our method experimentally on structural and functional (calcium dynamics) imaging of dozens of freely-moving \textit{C. elegans} across up to multiple square centimeter FOVs at micrometer resolutions at 120 fps.

\section{High-throughput microscopy with a diffractive pupil}
\subsection{Imaging system design}
Fundamentally, our microscope is a pupil-coded, nearly 4f imaging system, consisting of an objective and tube lens with a DOE placed in the pupil plane and a 2D array of sensors acting as a ``super sensor'' in the image plane (Fig. \ref{fig:teaser}). The purpose of the DOE is to create a sparse but extended PSF (see Sec. \ref{DOE} for design considerations) that enables not only filling of the gaps within the sensor array, but also extrapolation beyond the outermost sensors (Sec. \ref{fov_extrapolation}). The ``super sensor'' consists of a 6$\times$8 array of 48 8-bit monochrome sensors (3136$\times$4096 pixels, 1.1-\textmu m pixel pitch, 9-mm center-to-center spacing) for a total of 0.617 gigapixels (GPs) per snapshot, whose rectangular hull covers an area of 4.95 cm $\times$ 6.64 cm. The sensor array frame rate has two bottlenecks: the data transfer bandwidth of 6.17 GB/sec and the sensor readout rate. At full 0.617 GP sampling, the frame rate is bandwidth-limited to 10 fps, which in practice is too slow for imaging freely moving \textit{C. elegans} at high resolution. To achieve higher frame rates, we use 4$\times$4 pixel binning to achieve 120 fps (readout-limited). %In one of the imaging modes described below, we use 2$\times$2 binning across 6$\times$6 sensors to achieve 53 fps (also bandwidth-limited).
Since our method requires demixing overlapped images as part of an underdetermined system (we solve for more pixels than we measure), our samples should be relatively sparse (see Supplementary Fig. \ref{fig:L1_scan}). 

We report on two different imaging configurations: fluorescence and darkfield, both of which share the same tube lens (Jupiter-36B, f = 250 mm, \textit{f}/3.5) and DOE. In the darkfield imaging mode, we use an FJW Industries lens (f = 90 mm, \textit{f}/1.0) as the objective (for a total magnification of $\sim$2.78$\times$), four pseudo-collimated (Thorlabs ACL2520U-A, f = 20 mm) off-axis green LEDs (Thorlabs M530L4) as the illumination source, and a 514/2nm bandpass filter (Alluxa) in the pupil plane. The purpose of the narrowband filter is to reduce the radial spectral blurring of the nonzero diffraction orders, which would otherwise degrade spatial resolution (Supplementary Sec. \ref{spectral_blurring}). Since darkfield imaging has in principle zero background, it promotes sample sparsity. We also introduce a 40-mm-diameter aperture near the DOE, defining the pupil position and effective system resolution. 

In fluorescence imaging mode, we replace the objective with a Rodenstock lens (f = 42 mm, nominal NA = 0.7, design wavelength = 532 nm) as the objective (for a total magnification $\sim$6$\times$), a high-power blue LED as the widefield Köhler illumination source (Thorlabs SOLIS-470C), a 480/40nm excitation filter (Chroma), and a 530/55nm fluorescence emission filter (Edmund) directly above the DOE. Compared to the darkfield mode, the emission filter bandwidth is significantly broadened to collect as much fluorescence emission as possible. Here, we take advantage of the fact that the objective lens was designed for one particular wavelength, resulting in extreme axial chromatic aberrations (i.e., wavelength-dependent focal shift). Thus, the PSF induced by the DOE still has a sharp peak, albeit with heavy tails that contribute to out-of-focus background. However, we empirically found that the loss in SNR due to increased the background was offset by the increase in SNR from having a significantly broader emission filter bandwidth, given the weakness of fluorescence signals (Supplementary Sec. \ref{spectral_blurring}).

\subsection{Diffractive optical element design} \label{DOE}
As the pupil phase modulating element, the DOE directly determines the imaging system's PSF, which we optimized for filling in the inter-sensor gaps to enable a 2D, contiguous, full-field reconstruction. To this end, there were five chief criteria we optimized for when designing the PSF: 1) At every point in the sample FOV, the PSF will result in light falling on at least one sensor, 2) the PSF should have minimal translational ambiguity, 3) the PSF is as narrow in extent as possible to minimize spectral blurring (due to the wavelength dependence of diffraction), 4) the PSF is as sparse as possible to minimize the impact of read noise, and 5) the PSF is robust to scaling errors, which may arise in practice due to deviations in the lens focal lengths. The second criterion ensures maximization of the rank of the linear forward model, which can be efficiently evaluated using a triple correlation (Eq. \ref{triple_correlation}) between the PSF, itself, and an indicator function, $\mathit{Mask}(x,y)$, which is 1 where there is a pixel and 0 in the gaps of the sensor array plane. This triple correlation is a generalization of the typical PSF autocorrelation procedure for evaluating the performance of standard imaging systems with gap-free sensors. See Supplementary Sec. \ref{DOE_details} for details on PSF optimization.

Optimized based on these design criteria, our PSF consists of an asymmetric distribution of 15 diffraction-limited spots (Sec. \ref{final_psf_design}, Fig. \ref{table:psf}, Table \ref{table:psf}). This design enables a simple (windowed) convolution as our forward model, concentrates light into a sparse set of spots for mitigating loss in SNR, and achieves a spatial extent (6.8 mm $\times$ 6 mm) that is wider than the largest gap in the sensor array (5.55 mm $\times$ 4.49 mm) to enable gap filling. Another benefit of the extended PSF is that we can extrapolate beyond the rectangular hull of the sensor array (4.95 cm $\times$ 6.64 cm) by the width of the PSF (i.e., to 5.63 cm $\times$ 7.24 cm; see Sec. \ref{fov_extrapolation}). %Thus, there is a tradeoff between the extrapolated extent and loss of resolution due to spectral blurring.

Designing our DOE consisted of computing the pupil phase pattern, $\phi(x,y)$, that generates this multi-impulse PSF, within the constraints of our fabrication method, grayscale lithography. We used the Gerchberg-Saxton algorithm \cite{gerchberg1972practical}, modified to discretize the phase values to 64 levels at each iteration. We then converted $\phi(x,y)$ to height values, $h(x,y)$, based on the refractive index of the photoresist (S1813) at a design wavelength of $\lambda$ = 515 nm ($n$ = 1.6546):
\begin{equation}
    h(x,y)=\frac{\lambda\phi(x,y)}{2\pi(n-1)}.
\end{equation}
We fabricated a custom multi-level (64-level) DOE based on $h(x,y)$, with a maximum height variation of 0.787 \textmu m (corresponding to 2$\pi$ at $\lambda$ = 515 nm), a pixel size of 6 \textmu m, and a diameter of 39.8 mm (Sec. \ref{DOE_fabrication}).

%Note, however, that since our method aims to reconstruct >4$\times$ more pixels than measurements, the inverse problem is always underdetermined, regardless of the PSF. Nevertheless, there are several desirable properties for a PSF that maximizes the rank of the linear forward model . Our optimized PSF is a 15-point multi-impulse

% \begin{enumerate}
%     \item it is always visible across the entire FOV
%     \item has minimal translational ambiguity
%     \item (this list should be in supplement)
% \end{enumerate}

\begin{figure}[!hb]
    \centering
    \centerline{\includegraphics[width=.9\linewidth]{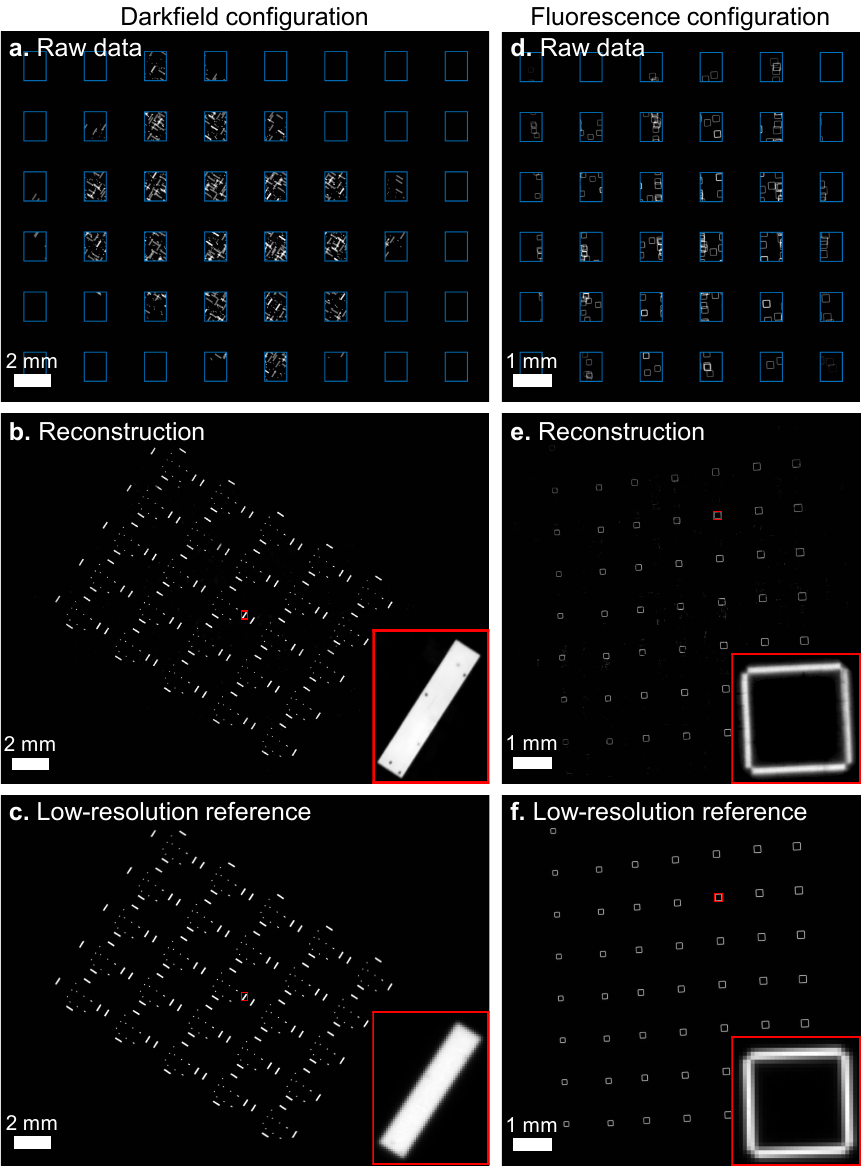}}
    \caption{Single-frame results using (a-b) darkfield configuration and (d-e) fluorescence configuration with a green LED, whose spectrum roughly matches that of green fluorescent protein (GFP). (a,d) Raw captured data of a photolithography mask. (b,e) Reconstructions of the full-scale image, with a zoom-in on the area denoted by the red box. (c,f) Low-resolution reference image, taken with a 1$\times$/NA=0.04 objective, with zoom-in on same area.}
    \label{fig:sample_recons}
\end{figure}

\subsection{Computational forward modeling and reconstruction}
\label{sec:forward_model}
In principle, the forward model describing our imaging system is simply a masked 2D convolution:
\begin{equation}\label{forward_model}
    P(\mathbf{r})=\left(\mathit{psf}(\mathbf{r})\otimes \mathit{Obj}(\mathbf{r})\right)\odot \mathit{Mask}(\mathbf{r}),
\end{equation}
\noindent where $\mathbf{r}=(x,y)$ is the 2D position, $P(\mathbf{r})$ is the forward prediction of the data, $\odot$ denotes elementwise multiplication, $\mathit{psf}(\mathbf{r})$ is the designed multi-impulse PSF, $\otimes$ denotes 2D convolution, $\mathit{Obj}(\mathbf{r})$ is the 2D object, and $\mathit{Mask}(\mathbf{r})$ is an indicator function defining whether there is a camera pixel at a given $\mathbf{r}$ position. An incoherent model is sufficient, with all quantities being real-valued. This is certainly true for fluorescence imaging, and approximately true for our darkfield imaging configuration, given the small coherence area of our LED illumination. Note that since the PSF is a collection of near-delta functions, Eq. \ref{forward_model} can also be interpreted as an incoherent superposition of multiple shifted copies of the object via the sifting property:
\begin{equation}
\begin{split}
P(\mathbf{r})&=\left(\sum_{i}A_i\delta(\mathbf{r}-\mathbf{\Delta r}_i)\otimes \mathit{Obj}(\mathbf{r})\right)\odot \mathit{Mask}(\mathbf{r})\\
 & = \left(\sum_{i} A_i\mathit{Obj}(\mathbf{r}-\mathbf{\Delta r}_i)\right)\odot \mathit{Mask}(\mathbf{r}),
\end{split}
\end{equation}
where $A_i$ and $\mathbf{\Delta r}_i=(\Delta x_i, \Delta y_i)$ are the amplitude and relative position of the $i^\mathit{th}$ impulse of the PSF, respectively.

In practice, the forward model is not shift-invariant due to distortions and aberrations in the objective and tube lenses. A common way to model distortion is with a radial distortion model, whereby the imaging magnification, $M(\mathbf{r})$, varies radially (e.g., barrel and pincushion distortion):
\begin{equation}
    M(\mathbf{r})=1+\sum_{i=1}^m a_i\left|\mathbf{r}-\mathbf{r}_0\right|^{2i}_2,
\end{equation}
which for simplicity is normalized by the nominal magnification, $M_0$, to 1 (i.e., so that the actual magnification is $M_0M(\mathbf{r})$), $\{a_i\}_{i=1}^m$ are the $m$ distortion coefficients, $\mathbf{r}_0=(x_0,y_0)$ is the center of distortion, and $|\cdot|_2$ is the L2 norm. This model assumes a rotationally symmetric imaging system; however, our microscope is not rotationally symmetric, even if all lens elements are perfectly aligned, since the DOE has an asymmetric diffraction pattern. Thus, we modeled distortions via two separable radial distortion models, one for the objective, one for the tube lens, 
\begin{equation}\label{separable_magnification}
\begin{split}
    M_\mathit{obj}(\mathbf{r})&=1+\sum_{i=1}^m a_i^\mathit{obj}\left|\mathbf{r}-\mathbf{r}_0^\mathit{obj}\right|^{2i}_2,\\
    M_\mathit{tube}(\mathbf{r})&=1+\sum_{i=1}^m a_i^\mathit{tube}\left|\mathbf{r}-\mathbf{r}_0^\mathit{tube}\right|^{2i}_2,
\end{split}
\end{equation}
so that each diffracted copy undergoes a unique distortion operation (Fig. \ref{fig:distortion_model}). For example, given a point in the object plane at position $\mathbf{r}_p$ and a 2D spatial shift $\Delta \mathbf{r}_i$ corresponding to the $i^\mathit{th}$ impulse of the PSF, the point experiences a magnification of
\begin{equation}\label{total_magnification}
    M(\mathbf{r}_p,\Delta \mathbf{r}_i)=M_0M_\mathit{obj}(\mathbf{r}_p)M_\mathit{tube}(-\mathbf{r}_p-\Delta \mathbf{r}_i).
\end{equation}

We can thus use these equations (Eq. \ref{separable_magnification}-\ref{total_magnification}) to generate the PSF at any desired position in the FOV for a fully shift-variant model. As such, we used direct convolutions, as opposed to FFT-based convolutions, which granted us a few more computational efficiencies. First, although the multi-impulse PSF has a large spatial extent, it only has 15 non-zero points contributing to the convolution. Second, we only needed to compute the convolution at spatial positions where there exists a sensor pixel (rather than for the full FOV, followed by masking). Thus, this shift-variant forward model can be written within the loss function as
\begin{equation}\label{loss}
\begin{gathered}
    \mathit{Loss} = \frac{1}{|S|}\sum_{\mathbf{r}\in S} \left(\left(\sum_i A_i \mathit{Obj}\left(M(\mathbf{r}_p,\Delta \mathbf{r}_i)(\mathbf{r}-\mathbf{\Delta r}_i)\right)\right)-\mathit{I(\mathbf{r})}\right)^2 + \lambda_1 \mathit{TV}(\mathit{Obj})+\lambda_2 |\mathit{Obj}|_1\\
    S=\{\mathbf{r} \mid \mathit{Mask}(\mathbf{r}) = 1\}
\end{gathered}
\end{equation}
where the first term is the mean square error (MSE), with the sum defined over positions in the image plane, $\mathit{I}(\mathbf{r})$, that contain detection pixels. We found that this fully shift-variant model was better than a patch-based shift-invariant model in terms of memory and accuracy, even when using FFT-based convolutions (Supplementary Sec. \ref{convolution_comparison}). The second term is an isotropic total variation (TV) regularizer and the third term is an L1 regularizer. We minimized Eq. \ref{loss} using (non-stochastic) gradient descent for each time point, also applying a non-negativity constraint at every iteration. Multi-sensor image data was acquired at 4$\times$ downsampling, and the associated gradient update operations could entirely fit on a 24-GB GPU and took <5 seconds to reconstruct each frame (i.e., 50 iterations at >10 iter/sec).
%subdividing the problem to fewer camera sensors when the full array did not fit in our GPU memory (24 or 48 GB). Specifically, image data acquired at 4$\times$ downsampling, and the associated gradient update operations, could entirely fit on a 24-GB GPU, and took 4-5 seconds per frame. %For image data acquired at 2$\times$ downsampling, we split the full 6$\times$8 array into two 6$\times$5 arrays, with overlapping images to minimize artifacts at the edges. In this situation, reconstructions took X seconds per frame.

% \subsection{Inverse optimization}
% To estimate the full-FOV object, $O(x,y)$, we minimized via gradient descent a loss function,
% \begin{equation}
%     L(O(x,y))=||P(x,y)-D(x,y)||^2
% \end{equation}

\section{Results}

\begin{figure}[!hb]
    \centering
    \centerline{\includegraphics[width=1.5\linewidth]{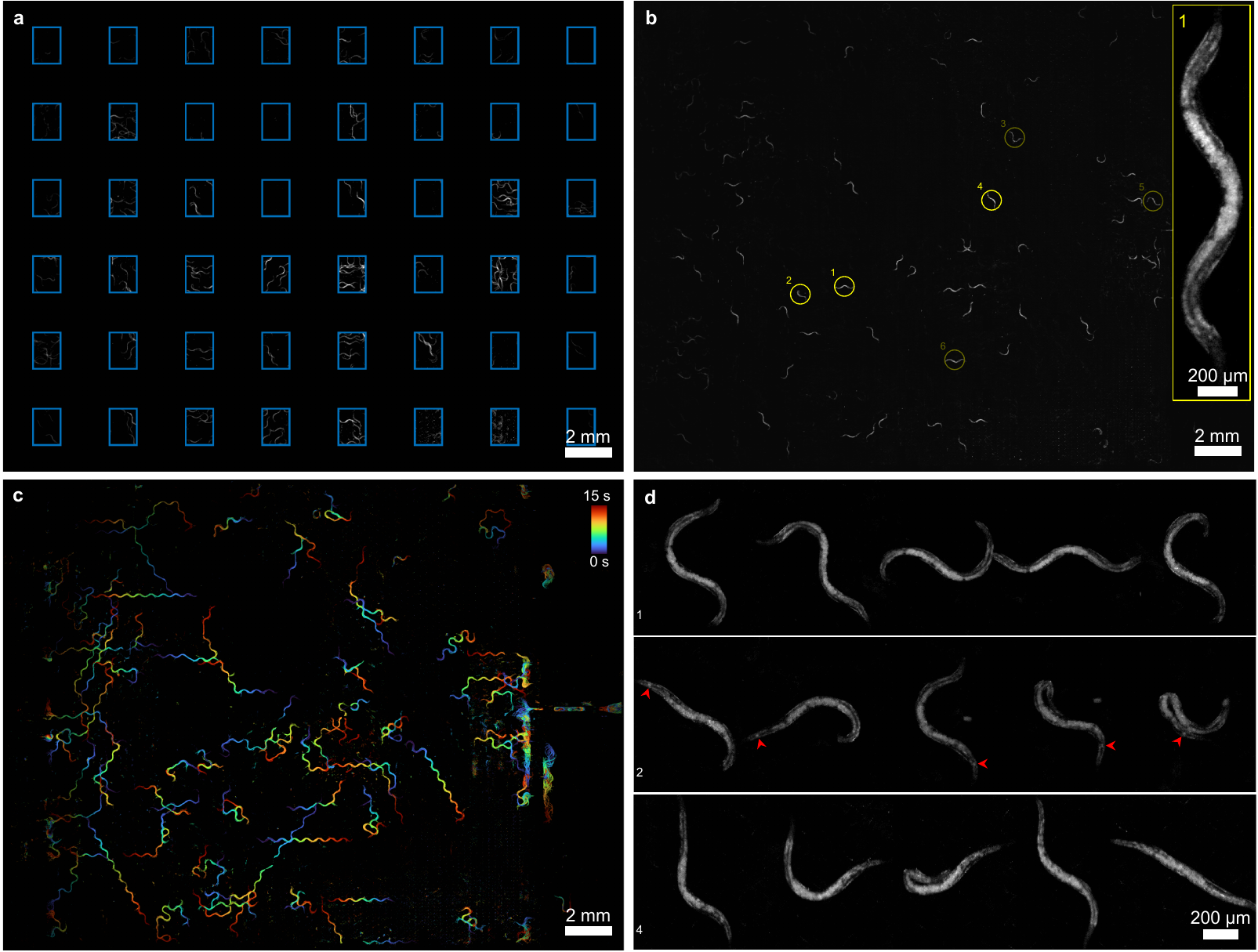}}
    \caption{High-speed (120 fps) darkfield imaging of wild-type \textit{C. elegans} at high resolution across a 20.0 $\times$ 26.4 mm\textsuperscript{2} FOV. (a) Raw data at a single time frame, displayed with each sensor in its correct relative position and with static background subtracted. The blue outlines indicate the sensor boundaries. (b) Reconstruction of the full-scale image from the data captured in (a). The circled and numbered worms were tracked in the full video. The inset shows a zoom-in of worm number 1. (c) The full video reconstruction represented with color coded by time. (d) Select frames showing the motion of a few of the worms. The frames are separated by 3.7 sec. The red arrowheads mark the posterior end of the gut, which can be tracked throughout the video. Please see Videos 1 and 2 for the full raw video and full video reconstructions, respectively.}
    \label{fig:darkfield}
\end{figure}

\subsection{System characterization}\label{system_characterization}
We characterized the spatially-varying resolution, field curvature, and depth of field (DOF) of both imaging modes by translating and acquiring z-stacks of a negative USAF target across the entire lateral FOV. From these image stacks, we computed an image sharpness metric based on the mean square image gradient to identify the most in-focus image across the extended FOV (Fig. \ref{fig:dof_curvature}), which are plotted in Figs. \ref{fig:usaf_darkfield_67}-\ref{fig:usaf_fluorescence_89}. Darkfield imaging mode can resolve down to group 8 element 4 (2.76 \textmu m full-pitch resolution) across most of the extended FOV, and at some positions, group 9. Thus, when we operate the sensors at 4$\times$ downsampling, the optical resolution and Nyquist sampling resolution (1.58 \textmu m pixel size at the sample) are nearly matched across the reconstructed $\sim$20.0 $\times$ 26.4 mm\textsuperscript{2} FOV, for a total per-frame pixel count of 12621 $\times$ 16655 = 210 MPs and a total imaging throughput of up to $\sim$25.2 billion pixels per sec. On the other hand, while the fluorescence imaging mode can resolve down to group 9 in the central cameras, it has significant aberrations in the periphery. However, the fluorescence imaging mode can resolve down to group 7 element 5 (4.38 \textmu m full-pitch resolution) across most of the 9.3 $\times$ 9.3-mm\textsuperscript{2} FOV. Thus, our fluorescence imaging resolution is aberration-limited, for a total imaging throughput of up to $\sim$2.17 GP/sec.

Plotting the best focus position and subtracting the plane of best fit (to account for tilt in the translational scan) reveal minimal field curvature across the FOV in both imaging modes (Fig. \ref{fig:dof_curvature}a,e). Finally, we computed the DOF across the extended FOV, defined as the axial full width half maximum (FWHM) of the sharpness metric (Fig. \ref{fig:dof_curvature}b,f). The DOF is larger further away from the center of the FOV, due to aberrations. Similarly, the fluorescence imaging mode has a larger DOF across the entire FOV, due to axial chromatic aberrations. Sample sharpness curves across all three spatial dimensions are shown in Fig. \ref{fig:dof_curvature}c,d,g,h.

%Resolution over FOV, field curvature, imaging throughput (pixel limit, aberration limit)

\subsection{Imaging of static samples}
As a proof of principle, we first imaged static photolithography masks. The raw, multiplexed data and the corresponding reconstructions are shown in Fig. \ref{fig:sample_recons} for both imaging configurations. It is apparent from the raw data that our multiplexed imaging design superimposes multiple copies of the sample onto the sensor array. The global features of our reconstructions closely match those of the low-resolution reference images, captured using a reference Nikon microscope with a low-magnification objective (1$\times$/NA=0.04), covering a FOV of 13.3 $\times$ 13.3 mm\textsuperscript{2} with a pixel size of 6.5 \textmu m. However, the zoom-ins show that our reconstructions have higher resolution (Fig. \ref{fig:sample_recons}b,c,e,f).

\subsection{Imaging of several dozen freely-moving \textit{C. elegans}}
To demonstrate the utility of our method in imaging highly dynamic samples, we imaged dozens of freely moving \textit{C. elegans} at high speed (120 fps) for 15 seconds using both imaging modes (1800 frames total). Using the darkfield mode, we performed structural imaging of wild-type \textit{C. elegans} freely moving across a 2 cm $\times$ 2 cm microfluidic device containing a hexagonal grid of cylindrical pillars, which mimic their natural soil environment (Fig. \ref{fig:darkfield}) \cite{Albrecht2011-hb, lockery2008artificial}. Prior to video reconstruction, we estimated and subtracted out the static background, removing darkfield signals from the pillars and any other static structures, leaving only the signal from the sparse distribution of dynamic \textit{C. elegans}. The full background-subtracted multi-sensor video data is shown in Video 1, a single frame of which is visualized in Fig. \ref{fig:darkfield}a. The multiplexed copies of the worms are computationally demixed, leading to a contiguous, gap-free video reconstruction of the worms and their trajectories across the 15-sec video (Video 2). A single frame of and a time-encoded image of the whole video are shown in Fig. \ref{fig:darkfield}b and c, respectively. The contiguity of the reconstruction across space and time enabled tracking of the worms, shown in the zoom-ins in Video 2 and Fig. \ref{fig:darkfield}d. Furthermore, the high resolution of our reconstruction enables tracking of structures within the worm throughout the 15-sec duration, such as the posterior end of the gut, circled in red in Video 2 and indicated by the red arrowheads in Fig. \ref{fig:darkfield}d.

\begin{figure}[!ht]
    \centering
    \centerline{\includegraphics[width=1.5\linewidth]{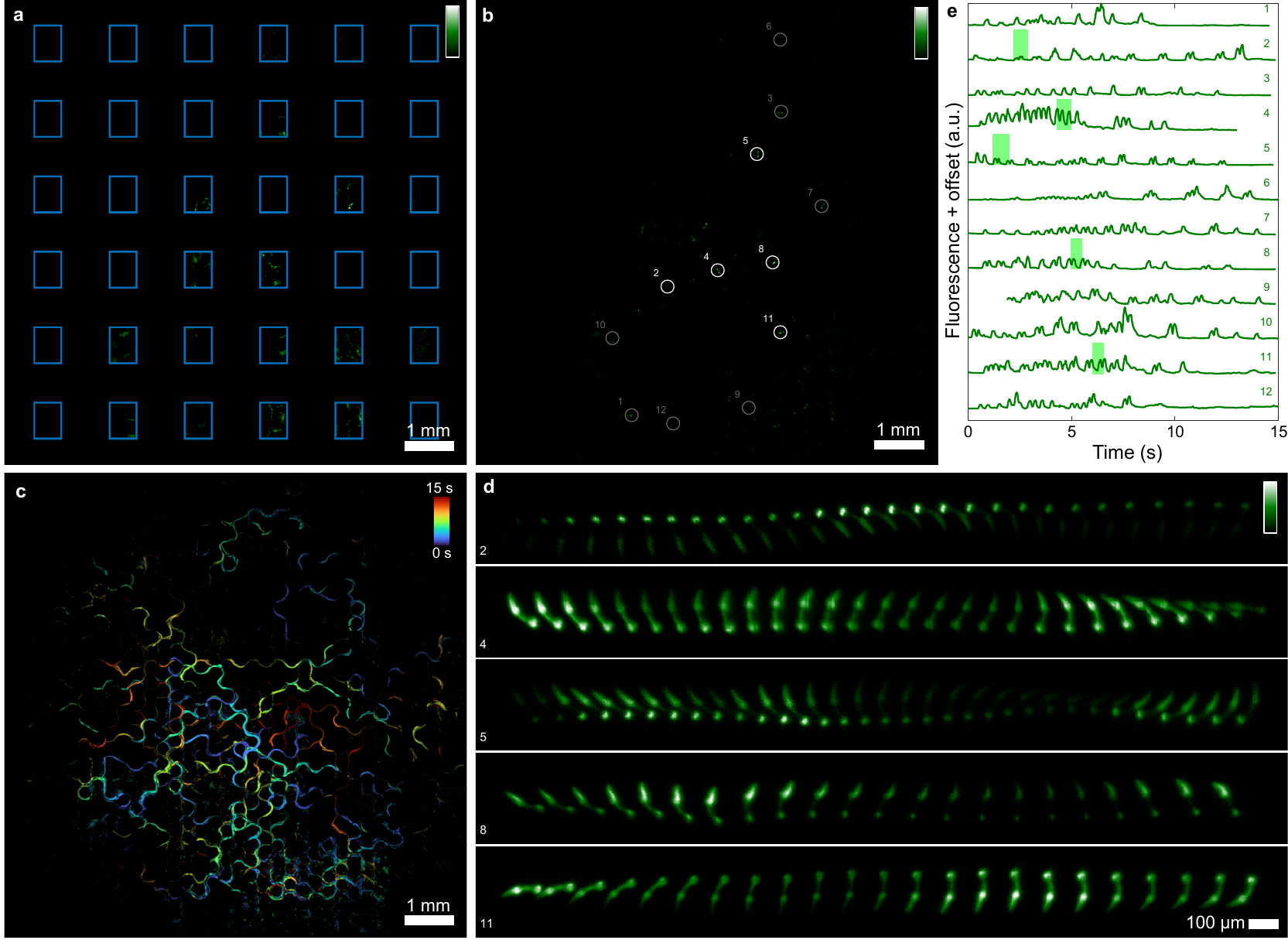}}
    \caption{High-speed  (120 fps) of \textit{C. elegans} expressing GCaMP8f in their pharynx. (a) Raw data (with static background subtraction) of a single frame, displayed with each sensor image in its correct relative position. The blue outlines indicate the sensor boundaries. See Video 3 for full video. (b) Reconstruction of the full-scale image from the timepoint in (a). The circled and numbered worms are analyzed in (d) and (e) as well as in Video 4. (c) The full 15-sec reconstructed video, color-coded by time. (d) Multiple frames (separated by 25 ms) showing the motion of 5 selected worms. (e) Fluorescence traces for the worms indicated in (b). The green highlighted time windows correspond to the image sequences plotted in (d). See Videos 3 and 4 for the full raw video and full video reconstructions and tracking of all 12 highlighted worms, respectively.}
    \label{fig:gcamp}
\end{figure}

Finally, using the fluorescence imaging mode, we imaged freely moving \textit{C. elegans} expressing GCaMP8f in the pharynx \cite{liu2024adapting}. Here, the GCaMP8f indicator responds to the influx of intracellular calcium ions during the \textit{C. elegans}' rapid pharyngeal pumping events (up to 4-5 Hz) as part of their feeding behavior. The worms were loaded in a smaller, 8 mm $\times$ 8 mm microfluidic chip with the same hexagonal array of pillars \cite{Larsch2013-av}. Video 3 shows the raw, background-subtracted, multi-sensor video data, a single frame of which is shown in Fig. \ref{fig:gcamp}a. The corresponding reconstructions are visualized in Video 4 and Fig. \ref{fig:gcamp}b,c. Note that both the raw data and reconstructions appear sparser than their darkfield counterparts Fig. \ref{fig:darkfield} because only the pharynx is fluorescently labeled. From the video reconstruction, we were able to track 12 individual worms throughout nearly the entire video and extract their pumping behavior from the fluorescence traces (Fig. \ref{fig:gcamp}e). A few sample pumping events are illustrated in Fig. \ref{fig:gcamp}d.

%\subsection{Calcium imaging of several dozen freely-moving \textit{C. elegans}}

\section{Discussion}
We present a very high-throughput computational microscope that employs a threefold strategy to expand FOV without sacrificing spatial resolution or frame rate: a discontiguous array of 48 independent camera sensors, diffractive pupil coding to further expand the FOV by >5.4$\times$, and a computational image reconstruction algorithm. Specifically, we designed and fabricated a DOE that, when placed in the pupil plane of our microscope, creates a multi-impulse PSF that encodes and compresses information in the gaps between the 48 sensors, enabling minimally ambiguous computational reconstruction. Our computational microscope, capable of achieving micrometer resolution over multiple square centimeter FOVs at 120 Hz frame rates, stands out among the highest throughput microscopes at 25.2 GP/sec. With these capabilities, we demonstrate high-resolution structural and functional imaging of dozens of \textit{C. elegans} in parallel.

%Our approach differs from other camera array-based microscopes

There are several avenues for future investigation that could further improve our microscope. We could extend our high-throughput 2D imaging system to 3D by modifying the DOE design to include depth encoding \cite{pavani2009three,linda2020fourier}. An important limitation of our design is that it requires lenses with relatively high SBPs, capable of achieving high resolution over a large FOV, which is practically difficult to achieve. As a result, our microscope objective and tube lens contain aberrations that prevent diffraction-limited resolution. In particular, given our pupil/DOE diameter of 40 mm and focal lengths of 90 mm (darkfield mode) and 42 mm (fluorescence mode), diffraction-limited performance would be 0.73 \textmu m (NA$\approx$0.43) and 1.45 \textmu m (NA$\approx$0.22), respectively. Moreover, fabricating an even larger DOE would further improve the theoretical diffraction-limited performance, though in practice would increase aberrations. Thus, calibrating the spatially-varying PSF and incorporating it into the forward model (Sec. \ref{forward_model}) could at least partially bridge the resolution gap. This could be particularly useful for the fluorescence imaging mode, whose PSF contains both geometric and chromatic aberrations due to the DOE and objective (Sec. \ref{spectral_blurring}). A hybrid diffractive-refractive design could also be employed to reduce these DOE-induced chromatic aberrations \cite{abrahamsson2013fast}. Careful PSF calibration with lensless approaches \cite{antipa2017diffusercam,zhao2020multi} could also help alleviate some of these challenges with aberration, albeit at the cost of SNR.

Despite these challenges, we have convincingly demonstrated experimentally the high-speed, high-resolution, wide-FOV imaging capabilities conferred by our microscope's highly parallelized and multiplexed design. With these new capabilities, our microscope can open up new applications not only in functional and behavioral imaging of large populations of freely moving model organisms, but also in diagnostic imaging for cell counting or detection of rare species, and in semiconductor inspection.

% extension to 3D
% DOE chromatic aberrations
% stop inside lens
% background subtraction
% aberrations ...

% limitations: darkfield illum from only 4 directions, lens aberrations, sensor speed, loss of light

\section{Methods}
\subsection{C. elegans preparation}
\textit{C. elegans} animals were maintained under standard conditions and fed OP50 bacteria \cite{Brenner1974-mt}. Wild-type worms were strain N2 (Bristol). The other strain used in functional imaging was INF418 \textit{nonEx106[myo-2p::GCaMP8f::unc-54 3'UTR]} \cite{liu2024adapting}.

Structural and functional imaging were performed using monolayer microfluidic devices \cite{Albrecht2011-hb,Larsch2013-av}, featuring arena areas 2 cm $\times$ 2 cm $\times$ 70 \textmu m or 8 mm $\times$ 8 mm $\times$ 70 \textmu m in size, respectively, where 200-\textmu m-diameter cylindrical pillars were arrayed hexagonally with a 300 \textmu m center-to-center spacing. In this structured environment, worms were inducted to crawl rather than swim \cite{Lockery2008-nn,Park2008-ys}. Before each experiment, ~150 young adult hermaphrodites were collected from cultivation plates using 2 ml of S-basal buffer (100 mM NaCl and 50 mM potassium phosphate; pH 6.0), washed twice in fresh 2 ml of S-basal buffer to remove excess bacteria, drawn into a loading tubing using a 1-ml syringe, and gently introduced into the microfluidic devices. These steps were completed within 10 min. Then, gravity-fed S-basal was allowed to flow through the arena for an additional 10 minutes, and the loaded worms were left to crawl freely for 1 hour before video recording began, allowing them to enter a starved state.

\subsection{DOE fabrication}\label{DOE_fabrication}
The diffractive optical element (DOE) was fabricated on a 50.8 mm diameter, 500 µm thick glass substrate using grayscale lithography. A layer of positive-tone photoresist (MICROPOSIT S1813 G2) was spin-coated at 1000 rpm for 60 seconds and baked on a hotplate at 110°C for 2 minutes. The design was patterned onto the sample with a laser pattern generator (DWL66+, Heidelberg Instruments GmbH). Post-exposure, the sample was heated at 50°C for one minute and developed in AZ Developer (1:1) for 1 minute and 20 seconds. A calibration sample, prepared and developed in the same way, was used to map photoresist depths to laser intensities from the pattern generator before the final write.
The DOE pattern consisted of a 6638 $\times$ 6638 matrix of heights, with each index measuring 6 µm x 6 µm. The maximum depth of the DOE pattern was 0.787 µm. A confocal microscope (Olympus LEXT OLS5000) was used to measure the heights of several outermost pixels in various locations on the sample. The average height difference between fabrication and ideal design was approximately 20 nm, with a standard deviation of 30 nm.

\subsection{Data preprocessing}
Prior to computational reconstruction (Sec. \ref{sec:forward_model}), we first performed two preprocessing steps. First, we estimated the background due to unrejected fluorescence excitation, darkfield signals from the pillars in the chips containing the \textit{C. elegans}, and any other signals that remain static throughout the video recording. To do this, for every pixel in the raw recordings, we computed the n\textsuperscript{th} percentile across time. We then subtracted this estimated background from each frame of the video. In the second step, we estimated the imaging system distortion and DOE orientation (Sec. \ref{sec:forward_model}) by jointly optimizing them with the reconstruction of a single frame from the video (or a superposition of multiple frames).

\subsection{Tracking}
Tracking of \textit{C. elegans} was performed almost entirely automatically using Trackpy in Fig. \ref{fig:darkfield} and Video 2. Automatic tracking of the GCaMP8f-expressing \textit{C. elegans}, however, was much more challenging, given the fluorescence signal fluctuations during pharyngeal pumping. Thus, we first performed automatic tracking using Trackpy, followed by manual tracking of the 12 worms highlighted in Fig. \ref{fig:gcamp} and Video 4.

\section*{Data availability}
% The raw video data for the \textit{C. elegans} videos in Fig. \ref{fig:darkfield}, Fig. \ref{fig:gcamp}, and Videos 1-4 will be made available.
The raw data underlying the results are currently not publicly available, but are available upon request.

\section*{Code availability}
The image reconstruction code will be made available on Github.

\section*{Acknowledgments}
We thank Monika Scholz for providing the GCaMP8f-expressing \textit{C. elegans} strain (INF418) and Leyla Kabuli for her constructive comments on the manuscript. This work was supported by the Office of Naval Research (N00014-22-1-2014), the Air Force Office of Scientific Research (AFOSR FA9550-22-1-0521), the National Institutes of Health (1RF1NS128772-01), and the Japan Society for Promotion of Science (JSPS KAKENHI 23KJ0349). This work was also supported in part by the Schmidt Science Fellows, in partnership with the Rhodes Trust.  Laura Waller is a Chan Zuckerberg Biohub SF investigator.

\section*{Contributions}
Conceptualization: KCZ, LW; Methodology: KCZ, CG, MI, NA, RM, RH, SK, LW; Software: KCZ, CG; Formal Analysis: KCZ, CG, MI; Investigation: KCZ, CG, MI, TH; Data Curation: KCZ, CG; Writing: KCZ, CG, MI, LW; Visualization: KCZ, CG, MI; Supervision: RM, SK, LW; Funding Acquisition: RM, SK, LW.

\section*{Disclosures}
KCZ was a consultant for Ramona Optics Inc. 
RH is a cofounder of Ramona Optics Inc., which is commercializing multi-camera array microscopes. RM is a co-founder of Oblate Optics, Inc that is commercializing DOEs. The remaining authors declare no conflict of interest.

\newpage
\renewcommand{\thesection}{S\arabic{section}}
\renewcommand{\thetable}{S\arabic{table}}
\renewcommand{\thefigure}{S\arabic{figure}}
\renewcommand\theequation{S\arabic{equation}}
\setcounter{figure}{0}
\setcounter{table}{0}
\setcounter{section}{0}
\setcounter{equation}{0}
\section*{Supplementary information}

\section{PSF design details}
\phantomsection  % this corrects error where refs to supp section link to main text instead
\label{DOE_details}
In this section, we explain in more detail the five criteria mentioned in the main text that we optimized for, followed by the exact optimization procedure. 

\subsection{Criteria for a good PSF}
\begin{figure}
    \centering
    \includegraphics[width=.8\linewidth]{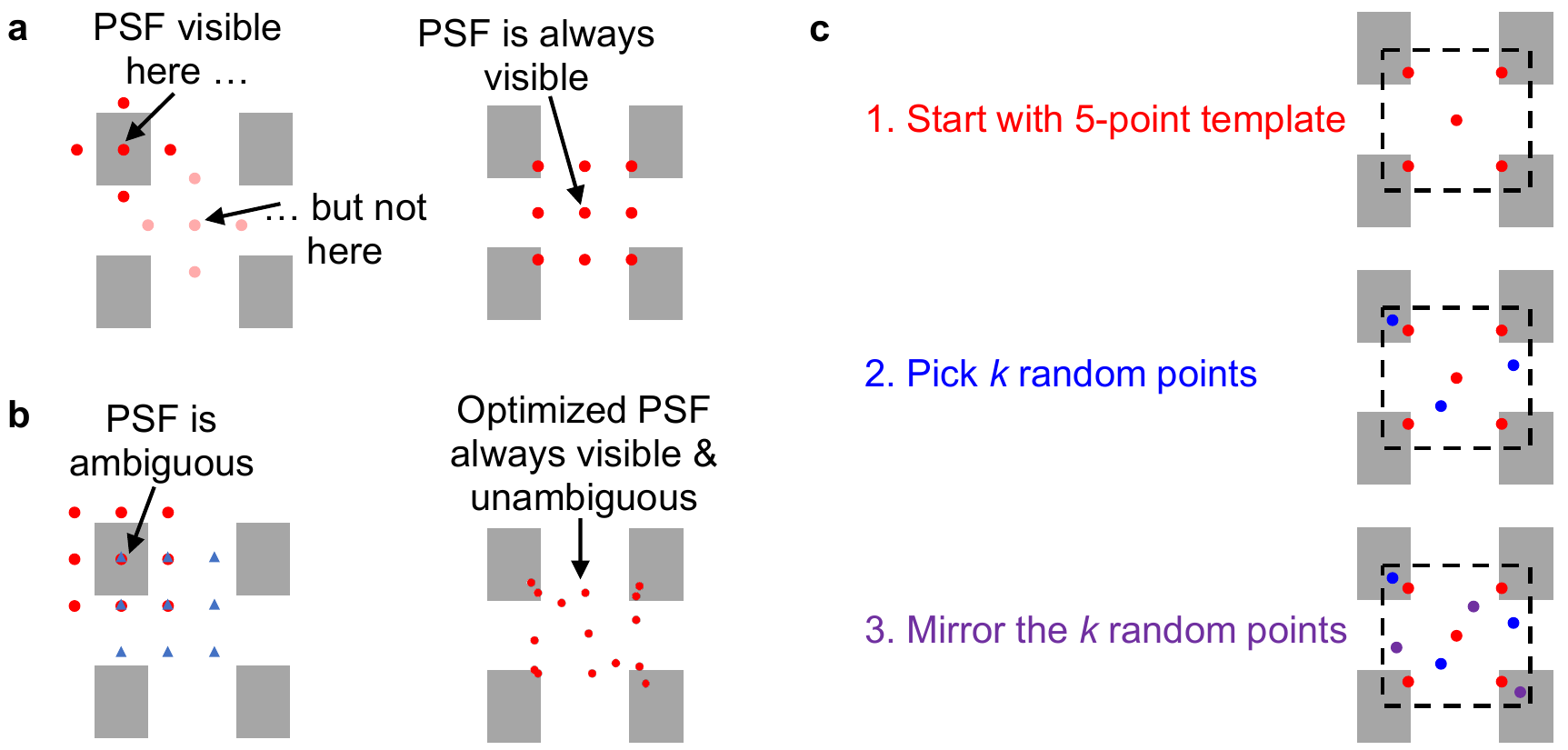}
    \caption{(a) The PSF should be visible on all sensors (gray rectangles), regardless of its position. While a 5-point PSF arranged in a ``+'' is not visible everywhere, a 3$\times$3 grid PSF slightly larger than the largest inter-sensor gap is visible everywhere.  (b) The PSF should be minimally ambiguous with translation. While the 3$\times$3 PSF is visible everywhere, it produces translational ambiguities. The optimized PSF has no strict ambiguities. (c) The three steps in proposing a candidate PSF in the PSF search algorithm.}
    \label{fig:psf_criteria}
\end{figure}

\phantomsection  % this corrects error where refs to supp section link to main text instead
\label{criteria}
\textbf{Pan-visibility.} First and foremost, the PSF should always be visible across the entire FOV on at least one sensor, including when the PSF is centered within a gap between sensors. This criterion ensures that we obtain a continuous, gap-free reconstruction. To fulfill this criterion, the PSF needs to be at least as large as the largest gap in the sensor array, which is 5.5504 mm by 4.4944 mm in the image plane. For example, a 3$\times$3 grid of dots with these dimensions satisfies this criterion (Fig. \ref{fig:psf_criteria}a); however, it has ambiguities (Fig. \ref{fig:psf_criteria}b) that renders reconstruction challenging, which leads us to the second criterion.

\noindent
\textbf{Minimal translational ambiguity.} Specifically, no two PSF positions in the FOV should give rise to the same measurement and they should be as dissimilar as possible (low autocorrelation). This argument is analogous to that of the standard, linear shift-invariant (LSI) imaging system (with no gaps), as minimizing PSF autocorrelation results in maximizing the frequency content of the MTF. Here, due to the discontinuous FOV, this criterion is harder to satisfy. For example, the aforementioned 3$\times$3 PSF fails this test (Fig. \ref{fig:psf_criteria}b), but would not fail if there were no gaps (although this PSF is autocorrelated). 

These first two criteria have intuitive interpretations, but could be understood more rigorously by examining the imaging problem more abstractly as a simple linear equation. To be concrete, let 
\begin{equation}
    \underset{(m \times 1)}{\mathbf{b}} = \underset{(m \times n)}{\mathbf{A}} \underset{(n \times 1)}{\mathbf{x}}
\end{equation}
describe a simplified forward model of our imaging system, ignoring distortion and aberrations, where $\mathbf{b}$ is a vector of length $m$, corresponding to the number of pixels across the sensor array, $\mathbf{A}$ is a Toeplitz matrix with the rows corresponding to shifted copies of the PSF, and $\mathbf{x}$ is a vector of length $n$, corresponding to the number pixels in the object reconstruction. Since there are gaps in the sensor array, $n>m$, so that this is an underdetermined problem. Further, since we're modeling an incoherent imaging system, the PSF and therefore the entries of $\mathbf{A}$ are strictly nonnegative. The best we can do for a general imaging system is maximize the rank of $\mathbf{A}$ (i.e., $m$); however, not all PSFs that maximize the rank of $\mathbf{A}$ are desirable. For example, a delta function PSF is the ``best'' PSF in terms of having zero translational ambiguity -- in other words, its corresponding $\mathbf{A}$ not only has maximal rank $m$, but also has orthogonal rows. However, such an $\mathbf{A}$ matrix is effectively an $m\times m$ identity matrix, as the remaining columns are all zeros -- in other words, certain parts of $\mathbf{x}$ would never be sampled, resulting in a gappy reconstruction. 

Thus, if we want $\mathbf{A}$ to be a nonnegative, $m\times n$ matrix without any zero-only columns, we must accept a PSF that has non-zero autocorrelation (i.e., the rows of $\mathbf{A}$ cannot be mutually orthogonal). Nevertheless, we want the rows and columns of $\mathbf{A}$ to be as close to orthogonal as possible -- doing so minimizes the condition number of $\mathbf{A}$. In compressive sensing, this is minimizing the mutual coherence of $\mathbf{A}$ (the max normalized correlation between any two columns of $\mathbf{A}$). Doing so ensures that no two positions in the FOV give rise to the same measurement on the sensor array. This minimization is also analogous to minimizing the autocorrelation of the PSF in a standard, gap-free LSI imaging system to improve its MTF, which we can generalize to our system by modulating the autocorrelation with the sensor array mask, $\mathit{Mask}(\mathbf{r})$. In particular, when testing candidate PSFs, rather than working with the very large $\mathbf{A}$ matrix, we can evaluate a normalized triple correlation:
\begin{equation}\label{triple_correlation}
    C(\mathbf{\Delta r}, \mathbf{\Delta r}')=\frac{\sum_r \mathit{psf}(\mathbf{r})\mathit{psf}(\mathbf{r}+\mathbf{\Delta r})\mathit{Mask}(\mathbf{r}+\mathbf{\Delta r}')}{\sum_r \mathit{psf}(\mathbf{r})^2\mathit{Mask}(\mathbf{r}+\mathbf{\Delta r}')},
\end{equation}
where we would like $C(\mathbf{\Delta r}, \mathbf{\Delta r}')$ to be close to 0 when $\mathbf{\Delta r}\neq(0,0)$. The reason why $\mathit{Mask}$ also needs to be cross-correlated is that the PSF should have low autocorrelation, regardless of whether it is centered in a gap or on a sensor. In practice, it is sufficient to evaluate Eq. \ref{triple_correlation} for $\mathit{Mask}$ corresponding to 3$\times$3 sensors, as our imaging system is ``periodically LSI''. 

For a gap-free imaging system with a PSF containing $k$ impulses, the lowest possible max autocorrelation for $\mathbf{\Delta r}\neq(0,0)$ would be $1/k$ -- at most one impulse overlapping, which is only achievable if the PSF doesn't have inversion symmetry (proof: PSF points $\mathbf{\Delta r}_i$ and $\mathbf{\Delta r}_j$ will simultaneously overlap with $-\mathbf{\Delta r}_j$ and $-\mathbf{\Delta r}_i$, respectively). Thus, our PSF design should not have inversion symmetry. However, for a gappy sensor, the triple correlation could be higher than $1/k$, because not all $k$ impulses may be visible at a given position in the sensor plane (this inflation is accounted for by the presence of $\mathit{Mask}$ in the denominator of Eq. \ref{triple_correlation}). 

\noindent
\textbf{Narrowest possible spatial extent.} While the PSF needs to be larger than the largest inter-sensor gap, it should be no larger due to the lateral chromatic aberrations of the DOE. In particular, since the diffraction pattern off of the DOE scales with the wavelength, an off-axis spot experiences radial smearing ($\delta r$) proportionally with distance from the zero diffraction order,
\begin{equation}
    \delta r\propto \frac{\Delta R\Delta\lambda}{\lambda},
\end{equation}
where $\Delta R$ is the radial distance from the zero order, $\Delta\lambda$ is the bandwidth of the detected light and $\lambda$ is the center wavelength (see Sec. \ref{spectral_blurring} for further discussions). Practically, this equation places a limit on the bandwidth of the detected light for a given spot size on the sensor. For the darkfield imaging mode, we use a 2-nm bandwidth bandpass filter as a compromise between resolution and light throughput. However, for the fluorescence imaging mode, we take advantage of the wavelength-dependent focusing properties of the objective (i.e., axial chromatic aberration) to mitigate the radial blurring effects. In practice, we specify a bounding box within which to propose the random points that make up the PSF.

\noindent
\textbf{PSF sparsity.} The fewer pixels into which our imaging system maps each point in the sample, the less sensor read noise we incur. This criterion also suggests that a good PSF design would be a collection of points. In practice, we have to balance this criterion with the minimal translational ambiguity criterion, which we found required a minimum number of points to satisfy. Furthermore, the more points, $k$, in the PSF, the smaller the max possible autocorrelation ($1/k$); however, we prioritized PSF sparsity due to the low light levels of our imaging applications.

\noindent
\textbf{Robustness to scaling errors.} Due to experimental imperfections, the actual PSF size may not match the design PSF size. For example, since the PSF size is proportional to the tube lens focal length, any error therein would lead to an artifactually large or small PSF. Thus, we searched for a PSF that would satisfy all the other criteria over a broad range of magnification errors.

%\subsection{Why should the PSF be a collection of points?}

\subsection{PSF optimization procedure}
The goal was to optimize a PSF parameterized by the positions of a distribution of points. Gradient-based methods are unsuitable because the gradients of the visibility and minimal ambiguity criteria with respect to the PSF point positions are zero almost everywhere. We thus opted for a gradient-free random search strategy, which involved systematically proposing random distribution points as the PSF candidates and checking adherence to the aforementioned criteria. In principle, there are an infinite number of suitable PSFs, so in practice, we ranked the suitable PSFs we obtained, based on their robustness to scaling errors. We caution that the PSF produced by the following procedure is not optimal by any metric, but rather is sufficient in practice for our imaging system.

\noindent
\textbf{Step 1: Propose candidate PSF.} We specified a number of points, $k$, the PSF can have, and a box bounding the PSF points, which we set to 1.3 times larger than the largest rectangular inter-sensor gap. We found that uniform random proposals overwhelmingly generated PSFs that failed the pan-visibility and minimal ambiguity criteria. Instead, we employed two strategies that increased the probability of generating a good PSF: 1) using a fixed five-point template just barely larger than the largest rectangular gap (Fig. \ref{fig:psf_criteria}c), and 2) proposing PSFs with inversion symmetry (we break the symmetry later -- see step 5 below). Due to this symmetry plus the zero order spot, $k$ is always odd.
%The basic PSF search strategy was to 1) randomly propose a candidate PSF with a fixed number of points, 2) check whether they satisfy the pan-visibility and minimal translational ambiguity criteria ($C(\mathbf{\Delta r}, \mathbf{\Delta r}')<1$ for all $\mathbf{\Delta r}\neq0 \text{ and } \mathbf{\Delta r}'$), and if so, 3) isotropically scale the PSF by a range of scale factors, and record the ratio between the max and min scale such that the pan-visibility and minimal translation ambiguity criteria are met, and finally 4)

\noindent
\textbf{Step 2: Check adherence to pan-visibility criterion}
This step was fast, and screened out many candidates, before computing the triple correlation in the next step, which was much slower.

\noindent
\textbf{Step 3: Compute triple correlation efficiently and identify shift ambiguities.} Since triple-correlating three 2D arrays (6D total) would be very slow, especially given the sizes of the arrays, we instead exploited a shortcut possible due to the multi-impulse parameterization of the PSF. Specifically, we considered Eq. \ref{triple_correlation} as a modulated cross-correlation between $\mathit{psf}(\mathbf{r})$ and a windowed version, $\mathit{psf}(\mathbf{r})\mathit{Mask}(\mathbf{r}+\mathbf{\Delta r}')$, where the former is a collection of $k$ points, $\mathbf{\Delta r}_i$, and the latter is a collection of $k'\leq k$ points, $\mathbf{\Delta r}_j$. Rather than create two 2D arrays to cross-correlate, we computed the pairwise 2D vectors, $\mathbf{\Delta\Delta r}_\mathit{ij}=\mathbf{\Delta r}_i-\mathbf{\Delta r}_j$, and counted the number of unique $\mathbf{\Delta\Delta r}_\mathit{ij}$ vectors. Note, then, that the computational complexity of the 2D cross-correlation scales like $k'k$ (where $k=15$ for our final PSF), rather than with the number of pixels in the extended PSF. 

After counting the unique $\mathbf{\Delta\Delta r}_\mathit{ij}$'s, we identified the maximum number of occurrences of any one $\mathbf{\Delta\Delta r}_\mathit{ij}$ (in other words, the number of points overlapping when the PSF and its modulated copy were offset by a shift of $\mathbf{\Delta\Delta r}_\mathit{ij}$) -- if this number was equal to $k'$ and there were at least two different $\mathbf{\Delta\Delta r}_\mathit{ij}$'s that occurred $k'$ times (one of them being $\mathbf{\Delta\Delta r}_\mathit{ij}=(0,0)$), then each additional $\mathbf{\Delta\Delta r}_\mathit{ij}$ besides $(0,0)$ was a potential 2D shift that produced an ambiguity in the proposed PSF. Put another way, this condition meant that all $k'$ points of the modulated PSF fully overlapped with the $k$-point PSF at at least two unique relative shifts (one of them being zero shift). We then tested for the only situation that this was not an ambiguity, which was if more than $k'$ points of the full PSF shifted by $\mathbf{\Delta\Delta r}_\mathit{ij}$ were seen by the sensor array -- the additional points break the ambiguity.

We must still repeat step 3 for all $\mathbf{\Delta r}'$ -- that is, shifting the PSF to different parts of the sensor array (which incidentally also may change $k'$).

\noindent
\textbf{Step 4: Save the PSF and compute its scale robustness.} Many candidate PSFs based on our sampling heuristic make it this far. We differentiate these candidate PSFs based on how robust they are to scaling errors. In particular, we isotropically scale the PSF by factors between $s=$ 0.8 and 1.2 and rerun steps 2 and 3. Supposing that scaling the PSF by any $s$ value within interval of $[s_\mathit{min}, s_\mathit{max}]$ passes steps 2 and 3, we report $s_\mathit{max}/s_\mathit{min}$ as the PSF's scale robustness factor. After running steps 1-3 many times and producing many candidate PSFs, we picked the one with the highest scale robustness factor.

\noindent
\textbf{Step 5: Break PSF symmetry.} We previously enforced inversion symmetry in the proposed PSFs to decrease the rejection rate. However, symmetric PSFs lead to larger autocorrelations, as discussed in Sec. \ref{criteria}. Essentially, if one picks any two non-opposite PSF points, $\mathbf{\Delta r}_i$ and $\mathbf{\Delta r}_j$, they will simultaneously overlap with $-\mathbf{\Delta r}_j$ and $-\mathbf{\Delta r}_i$, respectively, leading to autocorrelation peaks with amplitude $2/k$. There are $(k \text{ choose } 2) - (k-1)/2$ of such autocorrelation peaks (the second term arises due to the fact that picking opposite PSF points, $\mathbf{\Delta r}_i$ and $-\mathbf{\Delta r}_i$, does not lead to overlapping).

Thus, in this step, we perturb the PSF point coordinates to break the symmetry, allowing the max autocorrelation of the PSF to be $1/k$. In other words, we wanted to split the $2/k$-amplitude peaks into pairs of $1/k$-amplitude peaks, so that the split distance was as long as possible. If we pose this problem as a maximization of the minimum possible split distance among all the splits, it becomes equivalent to a circle packing problem (for $(k-1)/2$ circles). Since there's no guarantee that the perturbed PSF preserves the scale robustness factor, and since there's no inherent ordering of the $(k-1)/2$ circles, we randomly permuted the 2D perturbation vectors and picked the order that maximized the scale robustness factor.

\noindent
\textbf{Step 6: Repeat the above procedure for other values of $k$.} We wanted $k$ to be as small as possible to minimize the number of pixels into which the light would be distributed, but large enough that the PSF satisfies the pan-visibility and minimal ambiguity conditions with a large scale robustness factor. We empirically found that $k=13$ was sufficient, but yielded low scale robustness factors ($<$$1.05$), so we opted for $k=15$ to obtain larger robustness factors ($\sim$1.15).

\begin{figure}[!ht]
    \centering
    \includegraphics[width=0.4\linewidth]{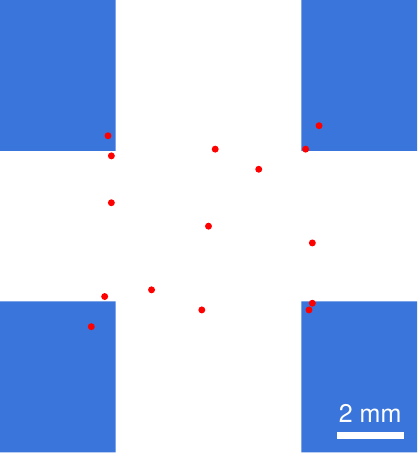}
    \caption{Optimized PSF, positioned such that the zero order is at the center of the gap of a 2$\times$2 sensor subarray. The individual PSF impulses are enlarged so that they are visible.}
    \label{fig:psf_on_2x2}
\end{figure}

\subsection{Final PSF design}
\phantomsection  % this corrects error where refs to supp section link to main text instead
\label{final_psf_design}
The final PSF design is shown in Fig. \ref{fig:psf_on_2x2} and its impulse coordinates are listed in Table \ref{table:psf}.

\begin{table}[]
    \centering
    \resizebox{\textwidth}{!}{%
    \begin{tabular}{*{15}{c|} c}
% x (mm) & 6.9 & 6.6 & 6.7 & 5.1 & 3.4 & 6.7 & 6.5 &  0.1 & 0.6 & 0.7 & 1.9 & 3.8 & 0.7 & 0.5 & 3.6 \\ \hline
%  6.2 & 0.7 & 2.7 & 4.9 & 0.7 & 0.9 & 5.5 & 0.2 & 5.9 & 3.9 & 1.3 & 5.5 & 5.3 & 1.1 & 3.2 

x (mm) &3.3&3.0&3.1&1.5&-0.2&3.1&2.9&-3.5&-3.0&-2.9&-1.7&0.2&-2.9&-3.1&0 \\ \hline

y (mm) &3.0&-2.5&-0.5&1.7&-2.5&-2.3&2.3&-3.0&2.7&0.7&-1.9&2.3&2.1&-2.1&0 \\

\end{tabular}
}
    \caption{The optimized PSF coordinates, relative to the zero order (the last coordinate).}
\label{table:psf}
\end{table}

% \section{A separable distortion model}
% Show example somewhere of shift-invariant model vs this model.

\section{Field of view extrapolation}
\begin{figure}[!ht]
    \centering
    \centerline{\includegraphics[width=1.2\linewidth]{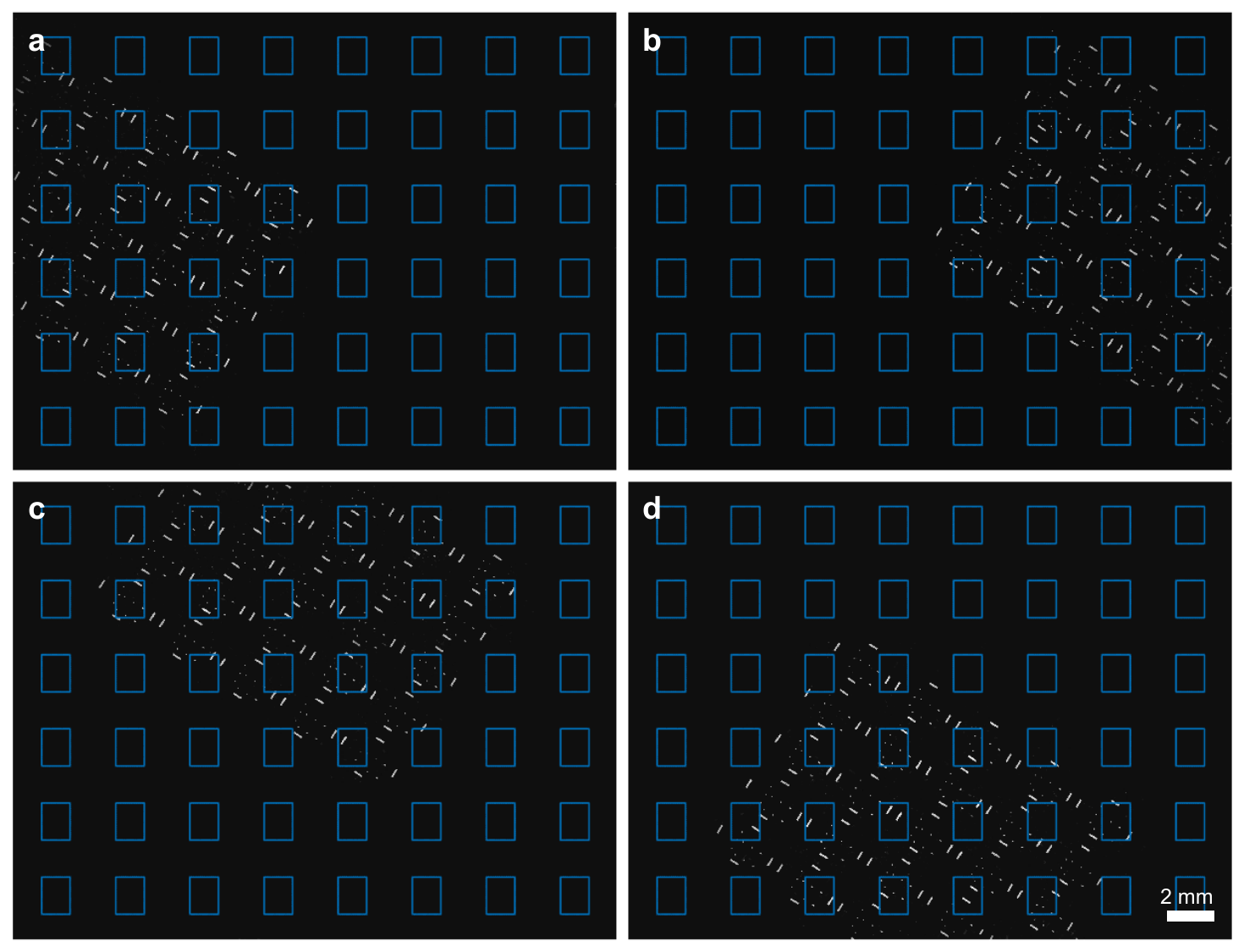}}
    \caption{Our method enables extrapolation beyond the rectangular hull of the sensor array (i.e., the smallest box bounding all 48 sensors, denoted by blue boxes). This figure shows the same target as in Fig. \ref{fig:sample_recons}b translated and reconstructed beyond the left (a), right (b), top (c), and bottom (d) edges.}
    \label{fig:extrapolation}
\end{figure}

\phantomsection  % this corrects error where refs to supp section link to main text instead
\label{fov_extrapolation}
The large spatial extent of the PSF on the sensor (6 mm $\times$ 6.8 mm) enables extrapolation beyond the native rectangular hull of the sensor array (4.95 cm $\times$ 6.64 cm). If we require half of the width or height of the PSF to be within the native rectangular hull, then the extrapolated FOV is simply the sum of the corresponding dimensions: 5.63 cm $\times$ 7.24 cm. Here, we validate that we can obtain good reconstruction quality in this extrapolation region by imaging a 2D target, translated well beyond the edge of the extrapolated FOV (Fig. \ref{fig:extrapolation}).

\section{Comparison of shift-invariant and shift-variant model}
\phantomsection  % this corrects error where refs to supp section link to main text instead
\label{convolution_comparison}

In the forward model described in Sec. \ref{sec:forward_model}, we introduced a convolutional forward model, interpreting it as an incoherent superposition of multiple shifted copies of the object without explicitly using convolution. This interpretation also allows full modeling of the distortion-induced spatial variance of the PSF. In this section, we justify this ``direct'' approach by demonstrating that it not only leads to more accurate reconstructions, but also uses the least memory and does not significantly trade-off speed compared to FFT-based approaches. In particular, here, we compare three algorithms: full-FFT, patch-FFT, and direct:

\noindent
\textbf{Full-FFT:} a shift-invariant model utilizing FFT-based convolution, as described in Eq. \ref{forward_model}. To achieve effective interpolation of the PSF's FFT, we pad the PSF to match the reconstruction size and then perform matrix multiplication in the frequency domain.

\noindent
\textbf{Patch-FFT:} a patch-wise shift-variant model using FFT-based convolution. The FOV is divided into patches corresponding to the sensor array (e.g., 6$\times$8 patches). For each sensor patch, we determined a support region encompassing all potential object-space points contributing to that sensor patch. The size of this region is the sum of the sensor size and the PSF size. In the forward model, we employ a sliding window to extract patches based on this support, apply FFT-based convolution with each patch's corresponding PSF, and crop the resulting sensor region to predict the measurement. 

\noindent
\textbf{Direct:} a fully shift-variant model utilizing direct superposition, as formulated in Eq. \ref{loss}. This approach implements a direct convolution, where the PSF is approximated as delta functions, with all zero-multiplication operations eliminated for computational efficiency.

\begin{figure}[!ht]
    \centering
    \centerline{\includegraphics[width=\linewidth]{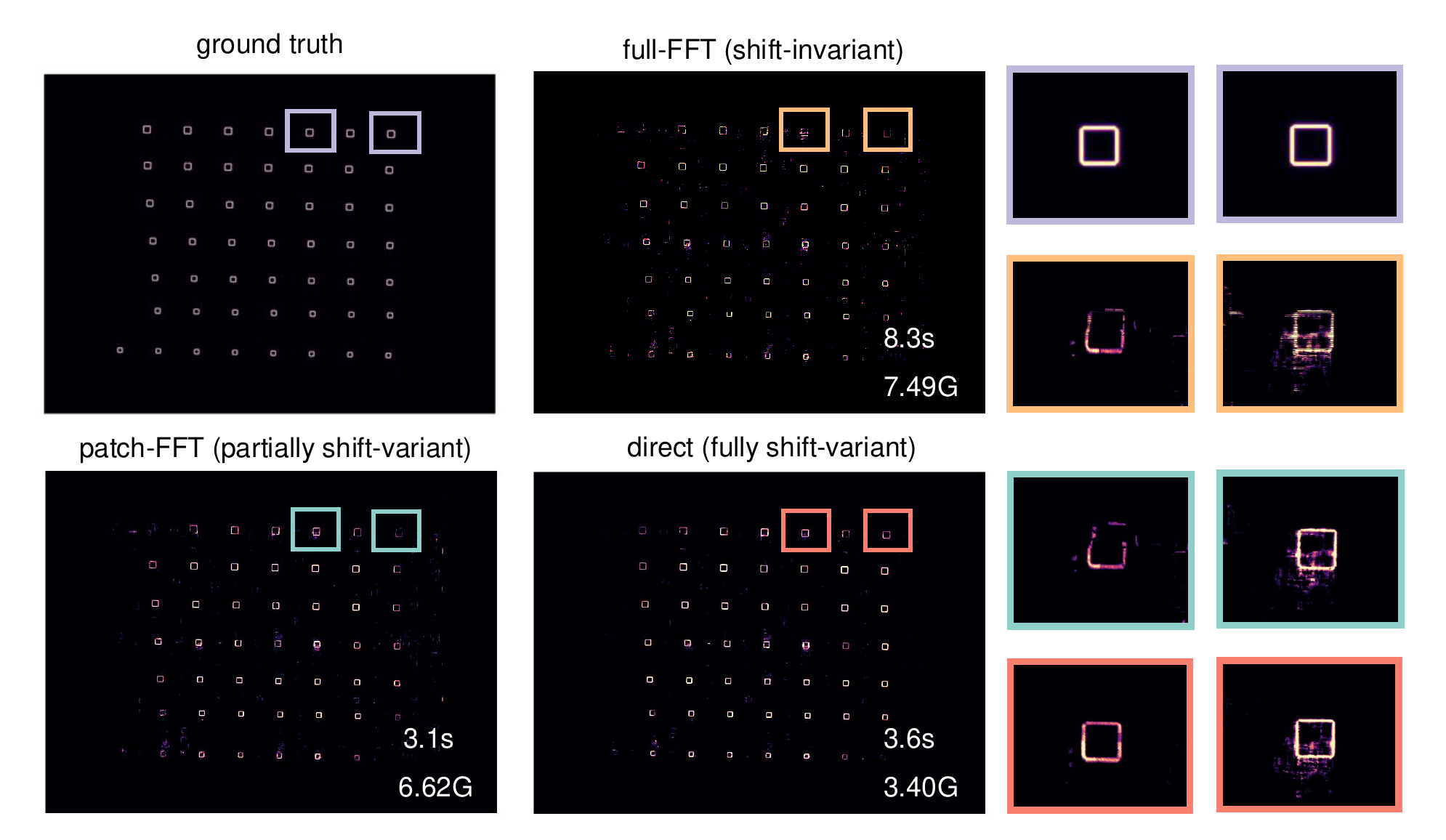}}
    \caption{Performance comparison across different algorithms: full-FFT, patch-FFT, and direct. Despite of being slightly slower than patch-FFT, the direct method achieves the best reconstruction quality, particularly in the peripheral regions. Additionally,  the direct method consumes the least memory, making it suitable for full-resolution reconstruction without downsampling.}
    \label{fig:alg_cmp}
\end{figure}

In Fig. \ref{fig:alg_cmp}, we use a lithography target as the sample and compare the algorithms' reconstruction against the ground truth captured by a regular microscope with 1x maginification. The algorithms' time and memory performance metrics are evaluated using the JAX framework with its just-in-time (JIT) compilation, executed on a single NVIDIA RTX 3090 GPU. We apply 8x downsampling to the measurement for all the methods, resulting in a reconstruction with dimensions of 6.3k$\times$8.3k (52 MP). The analysis includes two key performance indicators: the total computation time for 100 optimization iterations and the peak memory usage. These metrics are displayed in the bottom right corner of the figure, where the top row indicates the time consumption, and the bottom row shows the peak memory usage as reported by the memory\_stats field of the JAX device.

The direct approach yields the best reconstruction quality, followed by the patch-FFT and full-FFT approaches. When PSF shift variance is not properly accounted for (as in the full-FFT and patch-FFT method), the reconstruction suffers from either extra artifacts or eliminated objects. This trade-off underscores the importance of accurately modeling the shift-variant PSF distortion. 

We notice that the improvement in quality from patch-FFT over full-FFT is relatively minor. This is possibly because the patches are divided uniformly across the FOV, while the shift variance is not uniformly distributed for a lens system. It is optimized to have relatively small distortions in the central region, whereas the peripheral region exhibits more drastic PSF variations. We also observe that even with the direct method, the squares in the corner regions (especially in the bottom left corner) are not well reconstructed. This degradation primarily stems from lens vignetting, which naturally attenuates light intensity in peripheral regions. %The resulting reduced signal strength in these areas compromises the reconstruction fidelity.

To ensure a fair comparison, we generated the central and patch-wise PSFs for both FFT methods using lens parameters co-optimized according to Sec. \ref{sec:forward_model}. Specifically, we jointly optimize the 2D object $\mathit{Obj}(\mathrm{r})$ and the lens distortion parameters $\{a_i^\mathit{tube}\}_{i=1}^m, \{a_i^\mathit{obj}\}_{i=1}^m$ for both the tube lens and objective. The optimized lens parameters are then used to generate PSFs at the center of the full reconstruction and at each sensor's center for the full-FFT and patch-FFT methods, respectively. For the direct method evaluation, we disable lens parameter optimization since these parameters need only be co-optimized once.

% Additionally, the direct method's delta function approximation of PSF points enables optimization of PSF geometric distortion without adding extra computational overhead. The distortion parameterization is computationally efficient, requiring only $3\times 15=45$ parameters for the PSF dots' amplitude $A_i$ and relative positions $\mathbf{\Delta r}_i=(\Delta x_i, \Delta y_i)$, compared to the 16 million parameters needed for the full PSF with a 4k×4k support.

\section{Radial spectral blurring: SNR-resolution tradeoff}
\phantomsection  % this corrects error where refs to supp section link to main text instead
\label{spectral_blurring}
The wavelength dependence of DOEs can affect the spatial resolution of our computational microscope. In particular, the size of the PSF, as the far-field diffraction pattern of the DOE, is scaled in proportion to the wavelength:
\begin{equation}\label{psf_scaling}
    \mathit{psf(\mathbf{r}, \lambda)} = \mathit{psf}\left(\frac{\lambda}{\lambda_0}\mathbf{r}, \lambda_0\right),
\end{equation}
where $\mathbf{r}=(x,y)$ is the 2D spatial coordinate in the image plane, $\lambda_0$ is the design wavelength, and $\lambda$ is an arbitrary wavelength. The overall broadband PSF is thus
\begin{equation}\label{total_psf}
    \mathit{psf_{P(\lambda)}(\mathbf{r})} = \int_{0}^{\infty} P(\lambda)\mathit{psf(\mathbf{r}, \lambda)}d\lambda,
\end{equation}
where $P(\lambda)$ is the power spectral density of the detected light. As a result, the individual impulses of the PSF experience radial blurring, $\delta r$, in proportion to the bandwidth of the source, $\Delta\lambda$, and the radial distance from the zero order impulse,
\begin{equation}
    \delta r \propto \frac{\Delta\lambda}{\lambda_0}|\mathbf{r}|,
\end{equation}
suggesting that the PSF should be as small as possible and that the bandwidth of the detected light be as narrow as possible. The tolerance to bandwidth and PSF size depends on how $\delta r$ compares to the size of each PSF impulse, which in turn is determined by the tube lens NA and system aberrations. Assuming an impulse spot size of 8.8 \textmu m (the size that is critically sampled in the 4$\times$4 superpixel, with a pixel size of 1.1 \textmu m), a maximum radius of 3.5709 mm (given an inter-sensor gap size of 5.5504 mm by 4.4944 mm), and a design wavelength of 515 nm, the bandwidth that produces a radial blur of 8.8 \textmu m is $\Delta\lambda\approx$ 1.3 nm. Thus, we used a 2-nm bandpass filter for the darkfield imaging mode as a compromise between resolution and light throughput. %Fig. Xa,b shows a comparison of the USAF target imaged with the zero diffraction order vs. one of the first diffraction orders. 

For the fluorescence imaging mode, it would have been unacceptable to impose a 1-2-nm bandpass filter, given how weak fluorescence signals tend to be. Instead, we used a 55-nm bandpass filter and took advantage of the chromatic aberrations in the objective lens. Thus, when the wavelength deviates from the design wavelength, $\lambda_0$, we not only get scaling, but also defocus. We can modify Eq. \ref{psf_scaling} to incorporate blurring due to defocus:
\begin{equation}\label{psf_chromatic}
    \mathit{psf(\mathbf{r}, \lambda)} = \mathit{psf}\left(\frac{\lambda}{\lambda_0}\mathbf{r}, \lambda_0\right)\otimes\mathit{defocus}\left(\mathbf{r}, \frac{df}{d\lambda}(\lambda-\lambda_0) \right),
\end{equation}
where $\mathit{defocus}(\mathbf{r}, dz)$ is a 2D blur kernel due to a defocus of $dz$ and $df/d\lambda$ describes the focal shift per wavelength shift governing the axial chromatic aberration. 

Applying Eq. \ref{total_psf} results in a PSF that has a sharp in-focus component corresponding to the power at wavelengths close to the nominal wavelength, $\lambda_0$, and heavy tails due to the defocused components corresponding to the other wavelength components. With sufficient axial chromatic aberration, the lateral resolution is less impacted -- this principle has been previously applied for extended DOF imaging \cite{cossairt2010spectral}. However, there is still a tradeoff between total light throughput and background signal when spectrally filtering the fluorescence emission, with SNR being the ultimate metric of interest. That is, while increasing total light throughput improves SNR, increasing the background, although it can be digitally subtracted, introduces shot noise -- the total noise is a quadrature sum of the signal shot noise, the background shot noise, and the detector noise. Empirically, we found that the signal enhancement from increased light throughput outweighed the deleterious effects of background. %Fig. Xc,d compares the USAF target imaged using a zero vs. one of the first diffraction orders.

We have not included the effects of spectrally-dependent diffraction efficiency of the DOE in Eqs. \ref{psf_scaling}-\ref{psf_chromatic}, which increases the power in the zero order as the wavelength deviates from $\lambda_0$, modifying the effective $P(\lambda)$ for the nonzero orders. This effect could be compensated for in future designs by decreasing the zero order contribution at the $\lambda_0$, based on the expected fluorescence emission bandwidth. However, in practice, we found that the spectrally-dependent diffraction efficiency over the 55-nm bandwidth had minimal impact.

% The stronger the chromatic aberration, relative to the NA, the better the system lateral resolution

% \section{JAX vs Pytorch}

\section{L1 hyperparameter sweep}
\begin{figure}[!ht]
    \centering
    \centerline{\includegraphics[width=0.8\linewidth]{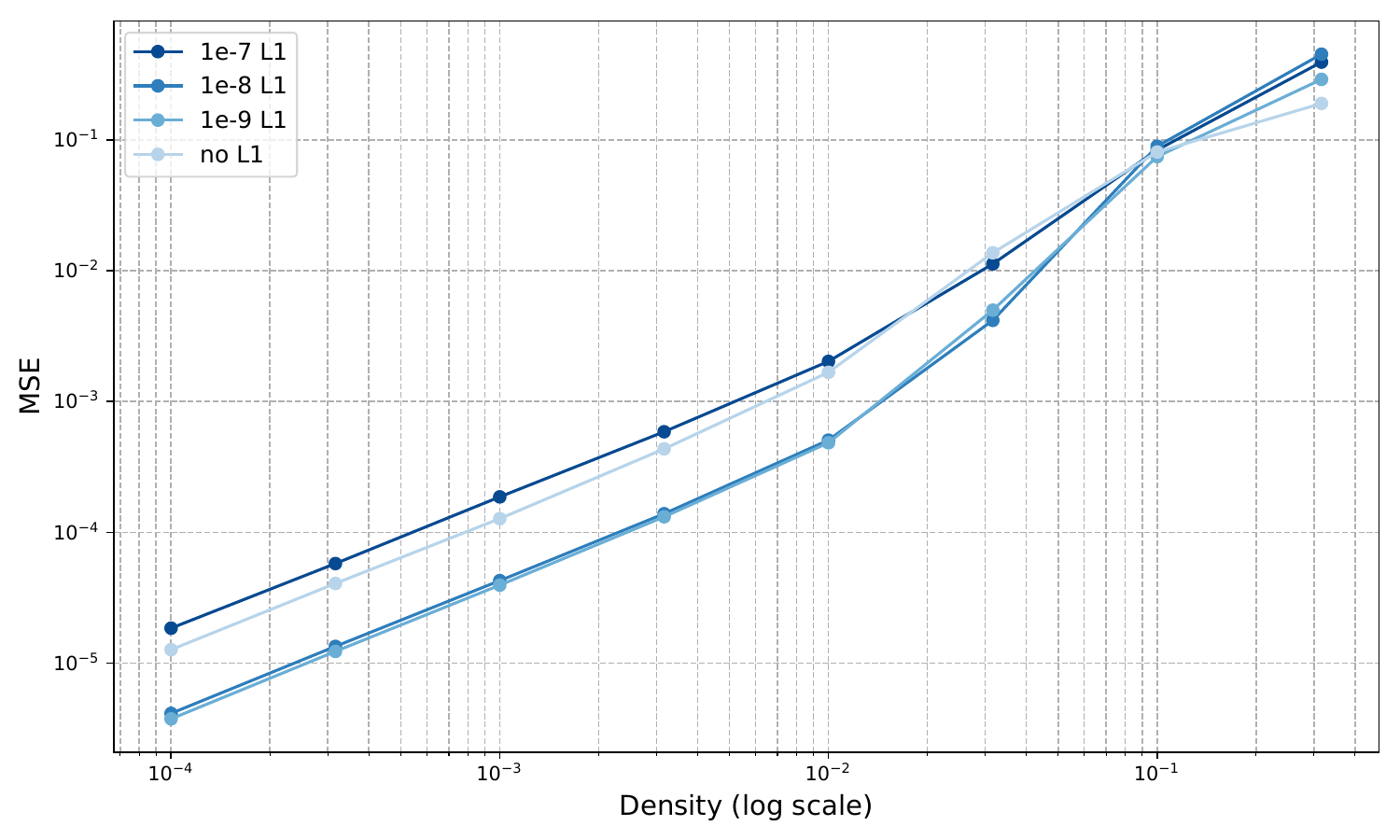}}
    \caption{Reconstruction MSE across different ground truth object density with different L1 regularization strength.}
    \label{fig:L1_scan}
\end{figure}
Fig. \ref{fig:L1_scan} demonstrates how reconstruction MSE varies with object density and L1 regularization strength. For dense objects, weaker sparsity constraints (lower L1 regularization) yield better results. In contrast, relatively sparse objects exhibit an optimal L1 regularization strength, with performance degrading when the regularization is either too weak or too strong.

These results were obtained through simulated measurements. We generated ground truth samples with randomly scattered dots at specified densities and simulate the measurements through the forward model. The analysis employed the direct method detailed in Sec. \ref{convolution_comparison}, assuming accurately known lens parameters.

% \section{Neural network stuff...}
% make summary slides

\section{Separable distortion model}
\begin{figure}
    \centering
    \centerline{\includegraphics[width=1.2\linewidth]{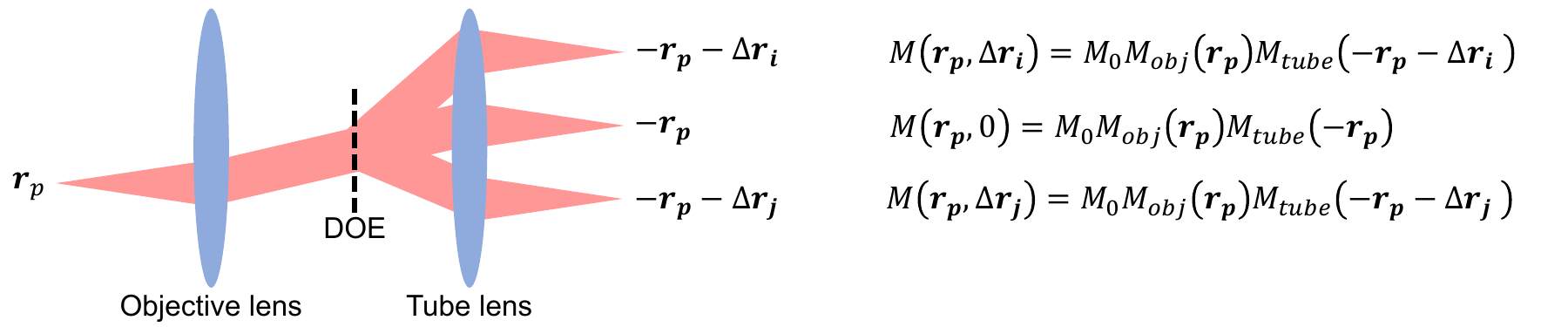}}
    \caption{We model the spatially-varying PSF of our imaging system based on a separable radial distortion model, whereby the objective and tube lens each has its own radial distortion model (i.e., radially-varying magnifications: $M_\mathit{obj}(\mathbf{r})$ and $M_\mathit{tube}(\mathbf{r})$). In an ideal imaging system with unit magnification, a position at point $\mathbf{r}_p$ in the object plane is mapped to a collection of points, $-\mathbf{r}_p-\Delta\mathbf{r}_i$, comprising the PSF. The total system magnification is thus a product of the independent contributions from the objective ($M_\mathit{obj}(\mathbf{r})$), the tube lens ($M_\mathit{tube}(\mathbf{r})$), and the nominal magnification ($M_0$, given by the ratio of the focal lengths of the objective and tube lens). }
    \label{fig:distortion_model}
\end{figure}

Fig. \ref{fig:distortion_model} illustrates the separable distortion model described in Sec. \ref{sec:forward_model}.

\section{Resolution, depth of field, and field curvature characterization}
\begin{figure}
    \centering
    \centerline{\includegraphics[width=1.3\linewidth]{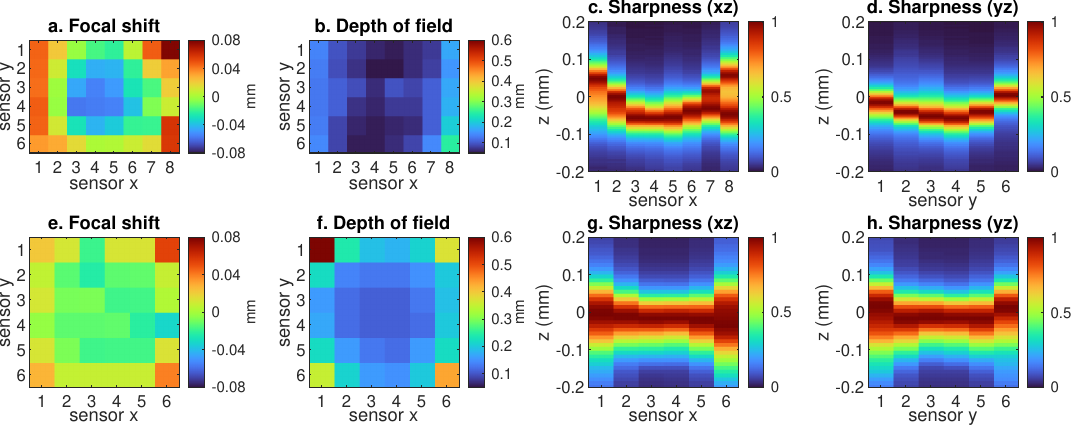}}
    \caption{Field curvature and depth of field (DOF) characterization for the darkfield (a-d) and fluorescence (e-h) imaging modes. (a,e) Focal shift at the centers of all 36 or 48 sensors. (b,f) Depth of field at the centers of all 36 or 48 sensors. (c,g) Sharpness as a function of depth and a central row of the sensor array, normalized to the max sharpness. (d,h) Normalized sharpness as a function of depth and central column of the sensor array.}
    \label{fig:dof_curvature}
\end{figure}

\begin{figure}
    \centering
    \centerline{\includegraphics[width=1.5\linewidth]{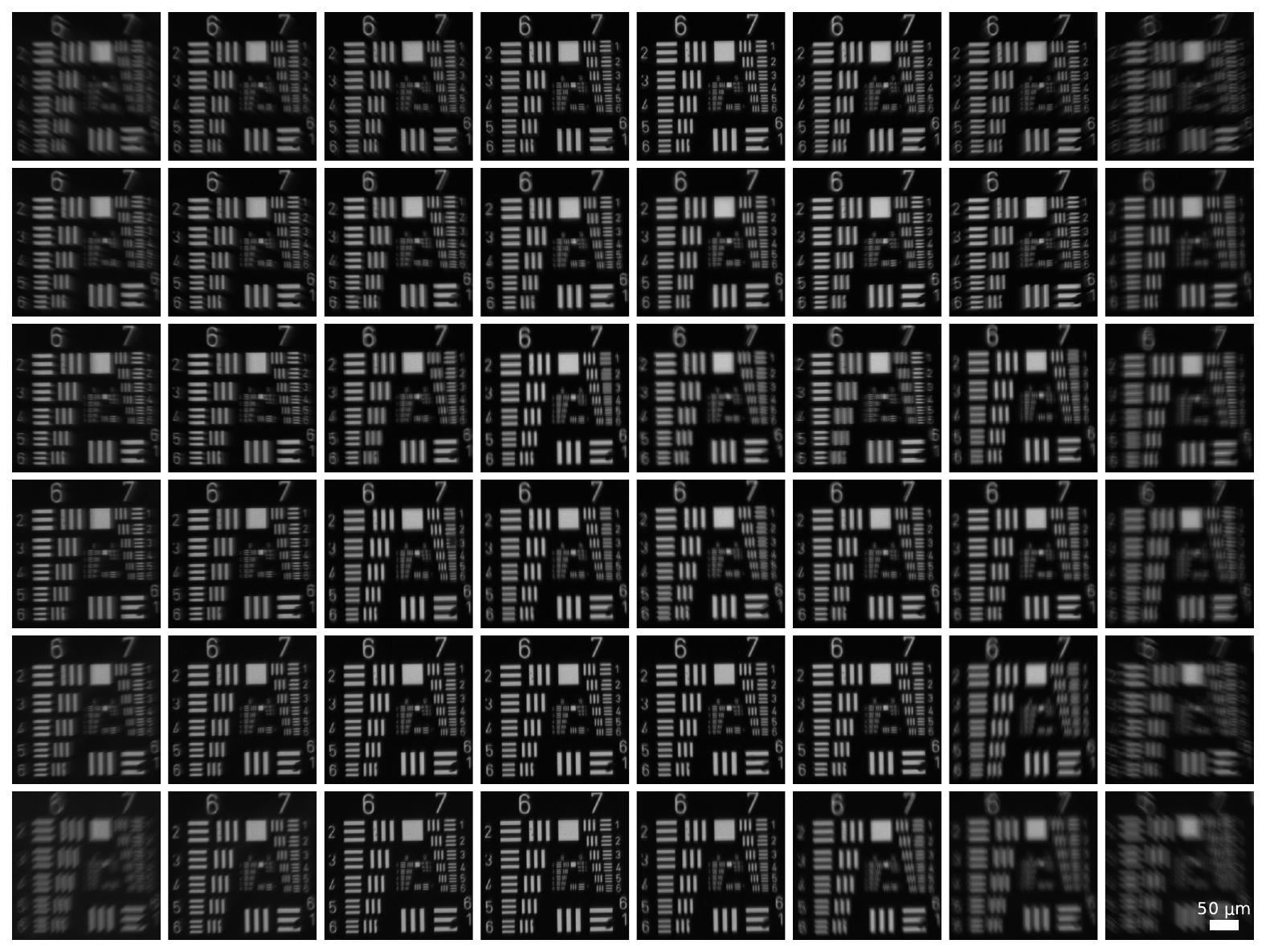}}
    \caption{Characterization of resolution across all $6\times8$ sensors in the darkfield imaging configuration. Groups 6 and higher are shown. Zoom-ins of groups 8 and 9 are shown in Fig. \ref{fig:usaf_darkfield_89}.}
    \label{fig:usaf_darkfield_67}
\end{figure}

\begin{figure}
    \centering
    \centerline{\includegraphics[width=1.5\linewidth]{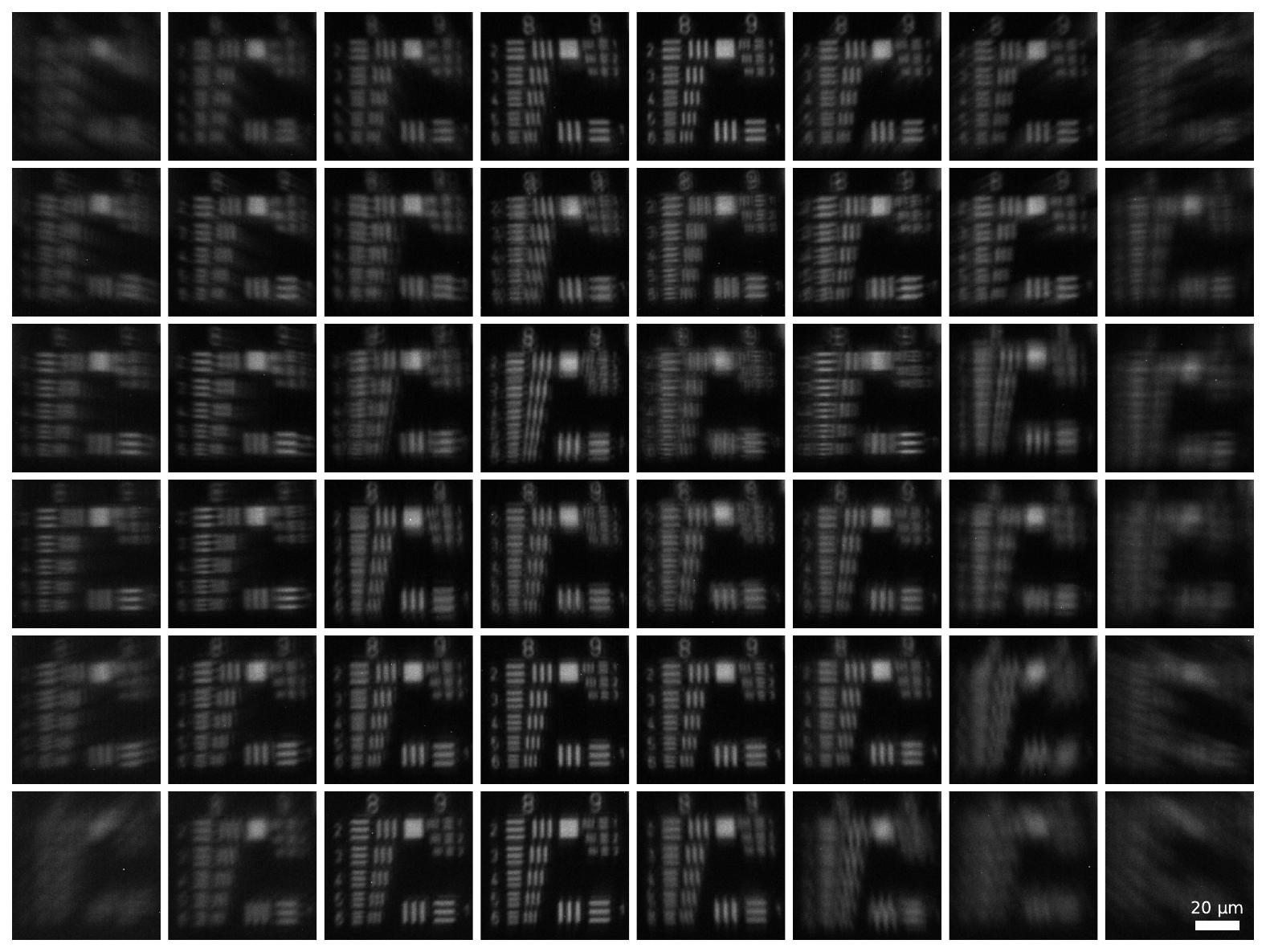}}
    \caption{Characterization of resolution across all $6\times8$ sensors in the darkfield imaging configuration. Groups 8 and 9 are shown.}
    \label{fig:usaf_darkfield_89}
\end{figure}

\begin{figure}
    \centering
    \centerline{\includegraphics[width=1.125\linewidth]{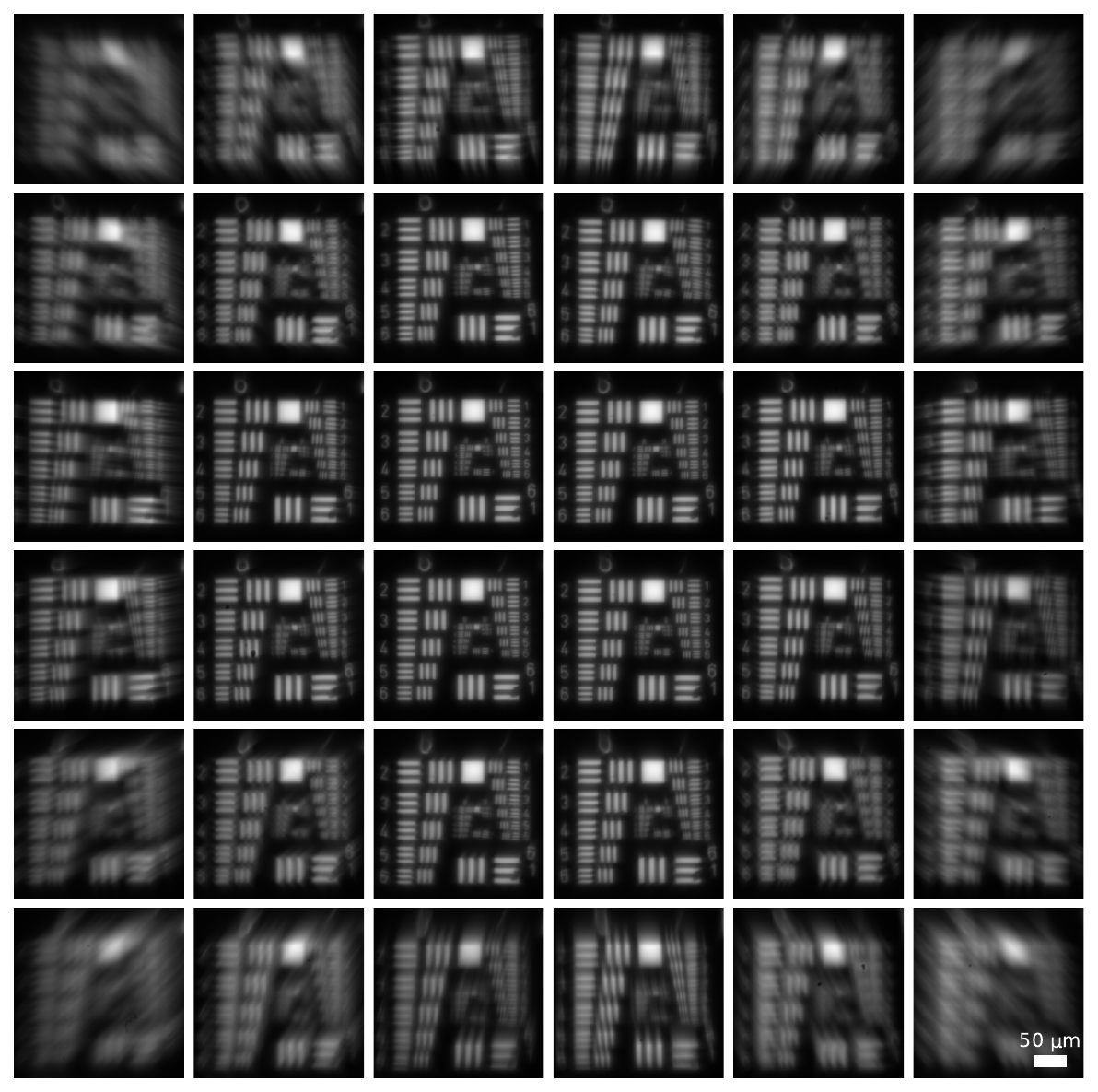}}
    \caption{Characterization of resolution across all $6\times6$ sensors in the fluorescence imaging configuration. Groups 6 and higher are shown. Zoom-ins of groups 8 and 9 are shown in Fig. \ref{fig:usaf_fluorescence_89}.}
    \label{fig:usaf_fluorescence_67}
\end{figure}

\begin{figure}
    \centering
    \centerline{\includegraphics[width=1.125\linewidth]{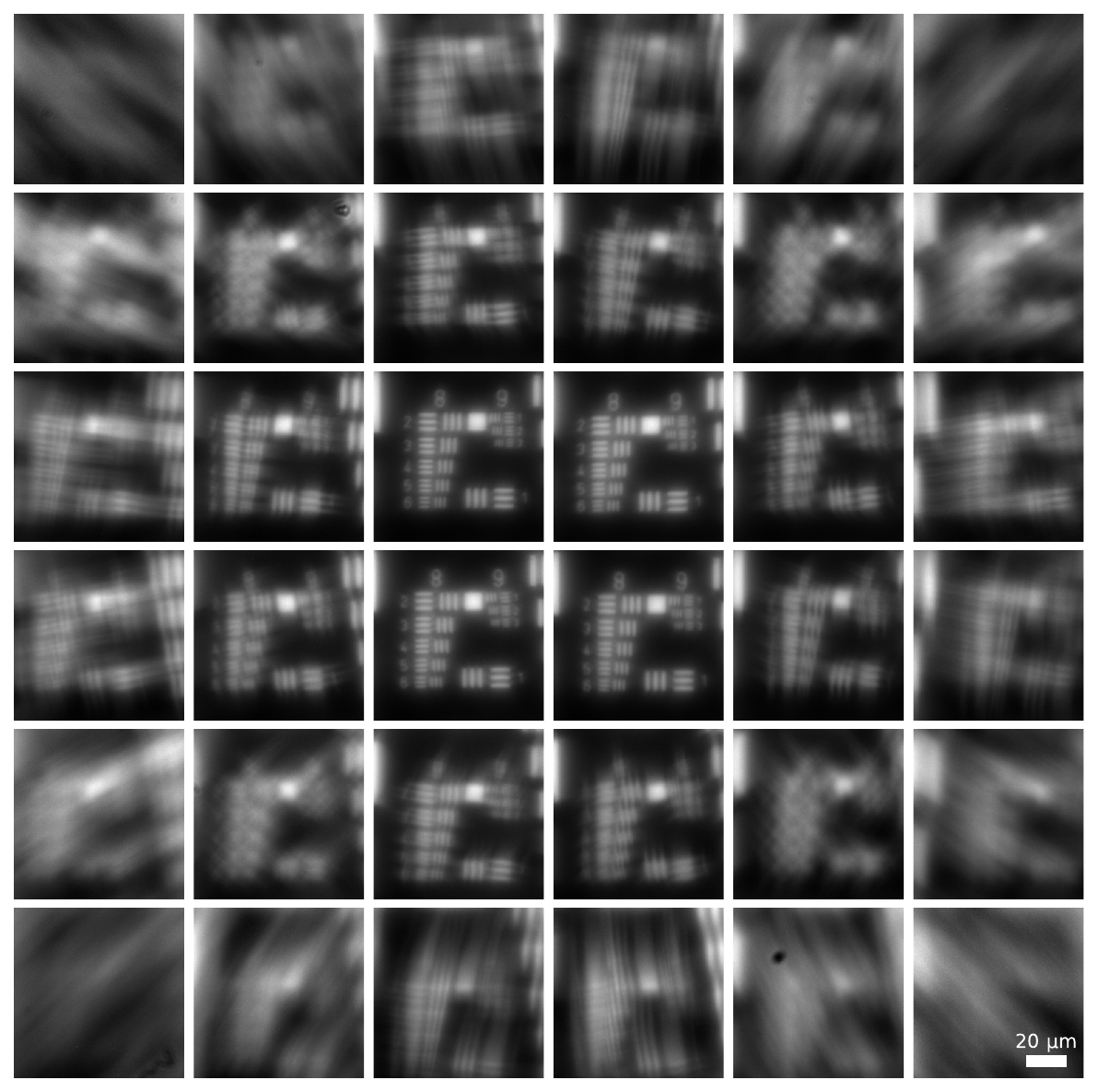}}
    \caption{Characterization of resolution across all $6\times6$ sensors in the fluorescence imaging configuration. Groups 8 and 9 are shown.}
    \label{fig:usaf_fluorescence_89}
\end{figure}

\phantomsection  % this corrects error where refs to supp section link to main text instead
\label{system_characterization+supp}
As described in Sec. \ref{system_characterization}, to characterize resolution, DOF, and field curvature of the two imaging configurations, we took z-stacks of a USAF target centered at each of the sensors (after magnification by the imaging system) and computed the sharpness based on the mean square image gradient (Fig. \ref{fig:dof_curvature}c,d,g,h). In most cases, we could identify the focal position via the argmax of this sharpness metric across z. However, recognizing that at certain field positions, our imaging system exhibits astigmatism, we identified the circle of least confusion as the focal position by finding the local minimum flanked by two local maxima, corresponding to the two astigmatic foci (e.g., see rightmost column of Fig. \ref{fig:dof_curvature}c). The USAF images at the best-focus positions across all sensors are plotted in Figs. \ref{fig:usaf_darkfield_67}-\ref{fig:usaf_fluorescence_89}. We removed any tilt between the lateral scan trajectory and the focal plane of the objective by fitting and subtracting the plane of best fit, yielding Fig. \ref{fig:dof_curvature}a,e. Finally, we computed the DOFs based on the FWHM of the sharpness curves (Fig. \ref{fig:dof_curvature}b,f).

\clearpage

\bibliography{sample}

\end{document}